\documentclass[aps,floats,prd,twocolumn,nofootinbib]{revtex4-2}

\usepackage{amssymb}
\usepackage{amsmath}

\usepackage{graphicx}
\usepackage{graphics}
\usepackage[caption=false]{subfig}
\usepackage{dcolumn}
\usepackage{color}

\usepackage{fancyhdr} 
\usepackage{hyperref}

\usepackage[utf8]{inputenc}
\DeclareUnicodeCharacter{00A0}{ }

 \makeatletter
 \renewcommand*{\fnum@figure}{{\normalfont\bfseries \figurename~\thefigure}}
 \renewcommand*{\@caption@fignum@sep}{\textbf{ : }}
 \makeatother

\makeatletter
\renewcommand*{\fnum@table}{{\normalfont\bfseries \tablename~\thetable}}
\makeatother

\def\VEV#1{\left\langle #1 \right\rangle}

    \newcommand{\be}{\begin{equation}}
  \newcommand{\ee}{\end{equation}}
    \newcommand{\ba}{\begin{align}}
  \newcommand{\ea}{\end{align}}

\newcommand{\Msun}{M_{\odot}}

\usepackage[dvipsnames]{xcolor}

\newcommand{\Mpcinv}{  \,\rm Mpc^{-1} }

\begin{document}

\title{Probing the Small-Scale Matter Power Spectrum with Large-Scale 21-cm Data}

\author{Julian B.~Mu\~noz\footnote{Electronic address: \tt julianmunoz@fas.harvard.edu}
} 
\affiliation{Department of Physics, Harvard University, 17 Oxford St., Cambridge, MA 02138}
\author{Cora Dvorkin\footnote{Electronic address: \tt cdvorkin@g.harvard.edu}} 
\affiliation{Department of Physics, Harvard University, 17 Oxford St., Cambridge, MA 02138}
\author{Francis-Yan Cyr-Racine\footnote{Electronic address: \tt 
fycr@unm.edu} }
\affiliation{Department of Physics and Astronomy, University of New Mexico, 270 Yale Blvd NE, Albuquerque, NM 87106}

\date{\today}

\begin{abstract}
The distribution of matter fluctuations in our universe is key for understanding the nature of dark matter and the physics of the early cosmos.
Different observables have been able to map this distribution at large scales, corresponding to wavenumbers $k\lesssim 10$ Mpc$^{-1}$, but smaller scales remain much less constrained.
In this work we study the sensitivity of upcoming measurements of the 21-cm line of neutral hydrogen to the small-scale matter power spectrum.
The 21-cm line is a promising tracer of early stellar formation, which took place in small haloes (with masses $M\sim 10^6-10^8M_\odot$), formed out of matter overdensities with wavenumbers as large as $k\approx100$ Mpc$^{-1}$.
Here we forecast how well both the 21-cm global signal, and its fluctuations, could probe the matter power spectrum during cosmic dawn ($z=12-25$).
In both cases we find that the long-wavelength modes (with $k\lesssim40$ Mpc$^{-1}$) are highly degenerate with astrophysical parameters, whereas the modes with $k= (40-80)$ Mpc$^{-1}$ are more readily observable.
This is further illustrated in terms of the principal components of the matter power spectrum, which peak at $k\sim 50$ Mpc$^{-1}$ both for a typical experiment measuring the 21-cm global signal and its fluctuations.
We find that, imposing broad priors on astrophysical parameters,
a global-signal experiment can measure the amplitude of the matter power spectrum integrated over  $k= (40-80)$ Mpc$^{-1}$ with a precision of tens of percent.
A fluctuation experiment, on the other hand, can constrain the power spectrum to a similar accuracy over both the $k=(40-60)$ Mpc$^{-1}$ and $(60-80)$ Mpc$^{-1}$ ranges even without astrophysical priors.
The constraints outlined in this work would be able to test the behavior of dark matter at the smallest scales yet measured, for instance probing warm-dark matter masses up to $m_{\rm WDM}=8$ keV for the global signal and $14$ keV for the 21-cm fluctuations. 
This could shed light on the nature of dark matter beyond the reach of other cosmic probes.
\end{abstract}

\maketitle

\section{Introduction}
\label{sec:intro}

The initial over- and under-densities of matter grew to give rise to the large-scale structure of our universe as we observe it today.
Different cosmic datasets have been able to probe the two-point function of these matter fluctuations---the power spectrum---across a broad range of scales.
Both the cosmic microwave background (CMB) and galaxy surveys serve as probes 
for comoving wavenumbers $k\sim (10^{-4}-0.1)\,\Mpcinv$~\cite{Gil-Marin:2015sqa,Abbott:2017wau,Aghanim:2018eyx}, with the Lyman-$\alpha$ forest 
and weak-lensing measurements reaching smaller scales, corresponding to $k \sim (1-10)\, \Mpcinv$~\cite{Abolfathi:2017vfu,Troxel:2017xyo}.
Nonetheless, higher wavenumbers are largely unconstrained.
Those hold information on the small-scale behavior of the cosmological dark matter (DM), as deviations from the standard cold-DM (CDM) paradigm leave an observable imprint onto the matter power spectrum~\cite{Spergel:1999mh,Weinberg:2013aya,Hui:2016ltb,Marsh:2015xka}.
Here we argue that 21-cm measurements during cosmic dawn are a powerful probe of the small-scale matter power spectrum, and thus of the nature of DM.

The cosmic-dawn era is expected to hold a trove of information about our cosmos (see, e.g., Refs.~\cite{Pritchard:2011xb,Furlanetto:2006jb} for reviews).
This epoch, halfway between the well-understood epoch of recombination and the local universe, is defined to begin when the first stars formed.
The ultraviolet radiation emitted by the stars coupled the hyperfine degrees of freedom of neutral hydrogen to its thermal state through the Wouthuysen-Field (WF) effect~\cite{Wout,Field,Hirata:2005mz}, allowing CMB photons with 21-cm wavelength to be resonantly absorbed by hydrogen in the intergalactic medium (IGM), producing a negative 21-cm temperature brightness.
Later on, the IGM is heated by the X-rays emitted by the first galaxies, producing observable 21-cm emission.

The combination of these two processes gives rise to the characteristic 21-cm absorption trough during cosmic dawn, and allows us to map early stellar formation through its effect on neutral hydrogen.
This will provide us with detailed knowledge of the astrophysics of the first galaxies, which are currently poorly understood~\cite{Bromm:2011cw}.
Moreover, by studying the timing of the 21-cm signal we can, therefore, reconstruct the evolution of stellar formation.
While many uncertainties remain about the properties of the first galaxies, we expect the first star-forming haloes to have masses $M_{\rm mol}\sim 10^6\,M_\odot$, large enough for gas to cool through molecular-hydrogen lines~\cite{Barkana:2000fd}.
Over time, however, it will become harder for those galaxies to form stars, as the accumulated Lyman-Werner (LW) photons (with energies in the $11.2-13.6$ eV band) dissociate hydrogen molecules, raising the minimum halo mass required for stellar formation~\cite{Machacek:2000us,Abel:2001pr,Haiman:2006si,Ahn:2008gh}.
This feedback process stops at the atomic-cooling threshold, with $M_{\rm atom} \sim 10^8\,M_\odot$, at which point the gas can cool down through atomic-hydrogen transitions~\cite{Oh:2001ex}.
All haloes below $M_{\rm atom}$ were formed out of extremely small-scale matter fluctuations, with comoving wavenumbers $k\gtrsim 40\Mpcinv$.
Suppressing said matter fluctuations delays the formation of structure, and thus all the 21-cm cosmic milestones.
Increasing them, on the other hand, will produce stellar formation at earlier times.
Here we propose using these measurements to indirectly probe the small-scale matter power spectrum.

Previous works have studied the effect of non-CDM models on the 21-cm signal~\cite{Lidz:2018fqo,Schneider:2018xba,Safarzadeh:2018hhg,Leo:2019gwh,Boyarsky:2019fgp,Lopez-Honorez:2018ipk,Yoshida:2003rm,Barkana:2001gr,Sitwell:2013fpa,Shimabukuro:2014ava,Bose:2016hlz,Lopez-Honorez:2016sur,Villanueva-Domingo:2017lae,Das:2017nub,Munoz:2018jwq,Escudero:2018thh,Mena:2019nhm,Yoshiura:2018zts}.
Those models typically suppress the matter power spectrum beyond some wavenumber $k$, delaying the onset of cosmic dawn.
Here, instead, we will follow a more generic approach, and study how well different types of 21-cm experiments can measure the small-scale matter power spectrum.
Our aim is twofold.
First, running the 21-cm simulations required to constrain each DM model can be costly, so a model-agnostic approach is preferred.
Second, we want to search for any deviation from the standard CDM scenario, regardless of its origin, which includes both dark-matter physics as well as non-standard inflationary scenarios~\cite{Kamionkowski:1999vp,Yoshiura:2019zxq}.

We will employ two observables: the 21-cm global signal and its fluctuations.
The global signal corresponds to the average absorption or emission across the entire sky.
While relatively inexpensive to perform, measurements of the global signal are difficult to interpret, as the cosmic component is buried on large foregrounds that have to be removed simultaneously from the data~\cite{Furlanetto:2006tf}.
The 21-cm fluctuations, on the other hand, are more costly to detect, requiring hundreds of antennae, although a prospective detection would be more robust, since foregrounds are limited to a ``wedge" of wavenumbers, with the rest of them expected to be foreground free~\cite{Liu:2009qga,Morales:2012kf,Datta:2010pk,Parsons:2012qh, Trott:2012md,Hazelton:2013xu,Pober:2013ig,Thyagarajan:2013eka,Liu:2014bba,Liu:2019awk,Liu:2014yxa,Thyagarajan:2015kla,Thyagarajan:2015ewa,Vedantham:2011mh}.
The dichotomy between these two probes makes them both useful observables.

We model the evolution of the 21-cm signal across cosmic dawn with a modified version of {\tt 21cmvFAST}~\cite{Munoz:2019rhi} (based on {\tt 21cmFAST}~\cite{Mesinger:2010ne,Mesinger:2007pd,Greig:2015qca}), in which we alter the input matter power spectrum at small scales.
In addition, we account for different astrophysical parameters that modulate the amount of stars formed, and their properties, which are presently unknown.
We fit for all parameters by performing a Fisher-matrix analysis, varying simultaneously the shape of the small-scale power spectrum, the astrophysical parameters, and---for the global signal---the nuisance foreground amplitudes.
For the global signal we consider an observatory based on the experiment to detect the global epoch-of-reionization signal (EDGES), whereas for the power spectrum we consider the hydrogen epoch of reionization array (HERA)\footnote{\url{https://reionization.org/}}.

We find that in the global signal there are nearly perfect degeneracies between the astrophysical parameters and the large-scale ($k\lesssim40\Mpcinv$) matter power, as they all rescale the amount of stars formed, whereas the power spectrum over scales $k=(40-80)\Mpcinv$ is less degenerate, and can be measured to $\sim 30\%$ precision.
The 21-cm fluctuations, on the other hand, have more discriminating power, due to the addition of spatial information.
This allows for degeneracies to be broken, and the matter power spectrum can be measured more precisely. 
In particular, we forecast that HERA can measure the matter power spectrum in the bins covering $k=(3-40)$, $k=(40-60)$ and $(60-80)\Mpcinv$, to 1.2\%, 16\%, and 30\% precision, respectively.

In order to overcome the large degeneracies between parameters, and between adjacent power-spectrum wavenumbers, we perform a principal component (PC) analysis of the matter power spectrum.
We find that the 21-cm line is most sensitive to wavenumbers $k=(40-80)\Mpcinv$, as those source the haloes that actively form stars during cosmic dawn.
We project different dark-matter models onto our PCs, forecasting that the 21-cm global signal can constrain warm dark-matter masses up to $m_{\rm WDM}=8$ keV, and the fluctuations up to $m_{\rm WDM} = 14$ keV.
This is to be compared with the current results from the Lyman-$\alpha$ forest, which constrain $m_{\rm WDM}>5$ keV~\cite{Irsic:2017ixq}, showing the promise of the 21-cm line.

This paper is structured as follows.
In Sec.~\ref{sec:model} we describe our model for the evolution of the 21-cm line and the simulations that we use.
Later, in Secs.~\ref{sec:global} and~\ref{sec:fluctuations} we show our results when considering upcoming measurements of the 21-cm global signal and fluctuations, respectively.
We use both datasets in Sec.~\ref{sec:PCs} to obtain model-agnostic PCs, before concluding in Sec.~\ref{sec:conclu}.

\section{The Model}
\label{sec:model}

We begin by describing our model to find the evolution of the 21-cm line across cosmic dawn, and how it is altered when the matter power spectrum is changed at different wavenumbers.
Throughout this work we assume fiducial cosmological parameters in agreement with the Planck 2018 data release~\cite{Aghanim:2018eyx}, of $\omega_b=0.0224$, $\omega_c=0.12$ $h=0.674$, with an amplitude $A_s=2.1\times 10^{-9}$ and tilt $n_s=0.965$ of primordial fluctuations.

\subsection{Collapsed Fraction}

Changing the matter power spectrum affects the amount of galaxies that are formed.
It does so indirectly, by altering the halo mass function (HMF) and the typical variance of overdensities of some mass.
We now explore each of those terms.

The main quantity that will determine the progress of 21-cm evolution in the universe is the collapsed fraction $F_{\rm coll}$ of baryons onto star-forming haloes (with mass $M>M_{\rm cool}$), defined as~\cite{Mesinger:2010ne,Tseliakhovich:2010yw,Munoz:2019rhi}
\be
F_{\rm coll} = \int_{M_{\rm cool}}^\infty dM M \dfrac{dn}{dM} \dfrac{f_g}{\rho_b} f_*(M),
\label{eq:Fcoll}
\ee
where $f_g$ is the fraction of gas collapsed into haloes of mass $M$, $\rho_b$ is the cosmic baryonic density,
and 
\be
f_*(M) = f_*^{(0)} \dfrac{\log(M/M_{\rm cool})}{\log(M_{\rm atom}/M_{\rm cool})}
\ee
is the fraction of gas that forms into stars, and it is set to a constant value $f_*^{(0)}=0.1$ for $M\geq M_{\rm atom}$~\cite{Fialkov:2012su}.
The HMF term above is given by
\be
\dfrac{dn}{dM} = \dfrac{\rho_M}{M^2} \left(- 2 \dfrac{d \log \sigma_M}{d \log M} \right ) \nu f(\nu),
\ee
where we use the Sheth-Tormen mass function~\cite{Sheth:1999su}, and
we remind the reader that $\nu = (\delta_{\rm crit}/\sigma_M)^2$.
Moreover, $\sigma_M$ denotes the variance of matter fluctuations on haloes of mass $M$, which is defined through
\be
\sigma_M^2 = \int \dfrac{d^3 k}{(2\pi)^3} |W_M(k)|^2 P_m(k),
\ee
where $W_M(k)$ is a window function, which we will set to be sharp in $k$~\cite{Schneider:2013ria}, as that better fits the collapsed fraction when considering models with suppressed power spectra~\cite{Schneider:2014rda}.
We note that sharp-$k$ window functions typically result in a smaller collapsed fraction at high $z$ than their top-hat counterparts, in better agreement with N-body simulations, as shown in Ref.~\cite{Schauer:2018iig}.

\subsubsection{Velocity-induced Acoustic Oscillations}

Throughout this work we will include the effects of the DM-baryon relative velocity on the formation of the first stars~\cite{Tseliakhovich:2010bj}.
By allowing baryons to stream away from DM minihaloes, large relative velocities suppress the gas fraction $f_g$ in Eq.~\eqref{eq:Fcoll}~\cite{Dalal:2010yt,Tseliakhovich:2010yw,Naoz:2012fr}, and increase the minimum mass $M_{\rm cool}$ that a halo needs to form stars~\cite{Fialkov:2012su,Greif:2011iv,Stacy:2010gg,Hirano:2017znw,1701.07031}.
More relevant to our work, the relative velocities produce a complex phase between the baryons and dark-matter fluctuations, suppressing the matter power spectrum at small scales ($k\sim 20-200 \Mpcinv$) in a spatially inhomogeneous way~\cite{Tseliakhovich:2010bj,Naoz:2011if}, which if unaccounted for would appear as a departure from CDM.
All these effects translate into velocity-induced acoustic oscillations (VAOs) in the 21-cm power spectrum~\cite{Visbal:2012aw,Munoz:2019rhi,Dalal:2010yt,Fialkov:2014wka}, which can be used as a standard ruler to the cosmic-dawn era~\cite{Munoz:2019fkt}.
For simplicity we will assume that the changes in the matter power spectrum that we study affect equally dark and baryonic matter, and thus do not change their relative velocities.

\subsubsection{Lyman-Werner Feedback}

The first stellar formation is expected to have taken place in small haloes, where the main coolants of gas were atomic or molecular transitions of hydrogen.
The former requires haloes larger than $M_{\rm atom} \approx 3\times 10^7 M_\odot$~\cite{Oh:2001ex}, whereas the latter only
$M_{\rm mol} \approx 3\times 10^5 M_\odot$~\cite{Barkana:2000fd}, and thus dominated the early phases of cosmic dawn.
Nonetheless, the UV emission from the first stars in the LW band (with photon energies $E=11.2-13.6$ eV), efficiently dissociates molecular hydrogen, halting stellar formation in the smallest haloes~\cite{Machacek:2000us,Abel:2001pr,Haiman:2006si}.
Let us now quantify this process of LW feedback.

Defining $F_{\rm LW}=4\pi J_{21}$, where $J_{21}$ is the LW flux in units of $10^{-21}$ erg s$^{-1}$ cm$^{-2}$ Hz$^{-1}$ sr$^{-1}$, we model the mass necessary to form stars as in~\cite{Fialkov:2012su},
\be
M_{\rm cool} (F_{\rm LW}) = M_{\rm cool}(0) (1+B F_{\rm LW}^{0.47}),
\label{eq:Mcoolfeedback}
\ee
where $M_{\rm cool}(0)$ is the threshold mass in the absence of LW photons (with typical values of $M_{\rm cool}(0)\sim 10^{5-6}\,M_\odot$, depending on $z$ and the DM-baryon relative velocity), and $B=7$ is the best-fit value from the simulations of Ref.~\cite{Machacek:2000us}, although given the large uncertainties in this process, we will marginalize over the amplitude $B$ of the feedback in our analysis.
Throughout this work we use the box-averaged $J_{21}$ flux from Ref.~\cite{Fialkov:2012su} (for reference, it corresponds to the case of regular feedback strength in Ref.~\cite{Munoz:2019rhi}).
Given this flux, and including the effect of DM-baryon relative velocities, by $z=25$ (when we realistically can start observing the 21-cm line), $M_{\rm mol} = 2\times 10^6\,M_\odot$ (an order of magnitude larger than the starting molecular-cooling haloes), and LW feedback is complete by $z\approx15$ (i.e., $M_{\rm cool}=M_{\rm atom}$).

Thus, between the turning on of Lyman-$\alpha$ coupling and the end of heating ($z\sim 12-25$) we expect to be able to probe the power spectrum over the $k=(40-100)\Mpcinv$ range that sources haloes with masses $M = (2\times 10^6 - 3 \times 10^7)\, M_\odot$.
We do not consider the era after X-ray heating is complete, where small haloes might be prevented to form stars by complex feedback processes~\cite{Barkana:1999apa,Ahn:2006qu,Mesinger:2008ze,Wise:2007nb}, and instead focus on the cosmic-dawn era, where feedback is dominated by LW photons.

\subsection{Binned Power Spectrum}

To explore variations around our fiducial CDM model, we divide the wavenumbers into $k$-bins, changing the amplitude $a_i$ of the matter power spectrum in each bin as a free parameter.
We define the fiducial amplitude of matter fluctuations as
\be
\Delta^2_{m,\rm fid}(k) =  \dfrac{k^3}{2\pi^2} P_{m,\rm fid}(k),
\ee
where $P_{m,\rm fid}$ is the matter power spectrum, obtained from CLASS~\cite{Blas:2011rf}.
We vary this quantity as
\be
\Delta^2_m(k,\mathbf a) = \Delta^2_{m,\rm fid}(k)  \sum_i a_i f_i(k),
\label{eq:Pmatter}
\ee
where we choose the basis functions $f_i(k)$ to be log-space top-hats around consecutive wavenumbers $k_i$, and group the amplitudes---all with a fiducial value of unity---into a vector $\mathbf a=\{a_i\}$.
We divide our $k$-range, spanning from $k=20$ Mpc$^{-1}$ to  $k=108$ Mpc$^{-1}$, in 40 logarithmically spaced bins.
The lower end is chosen to be safely below the atomic-cooling threshold, whereas we have tested that we do not have sensitivity to modes with $k\gtrsim 100\Mpcinv$, as those correspond to haloes with masses $M\lesssim 10^6 M_\odot$, which are only above the molecular-cooling threshold for $z>25$~\cite{Visbal:2014fta}, where the noise is too large to reliable detect 21-cm.

\begin{figure}[hbtp!]
	\centering
	\includegraphics[width=0.5\textwidth]{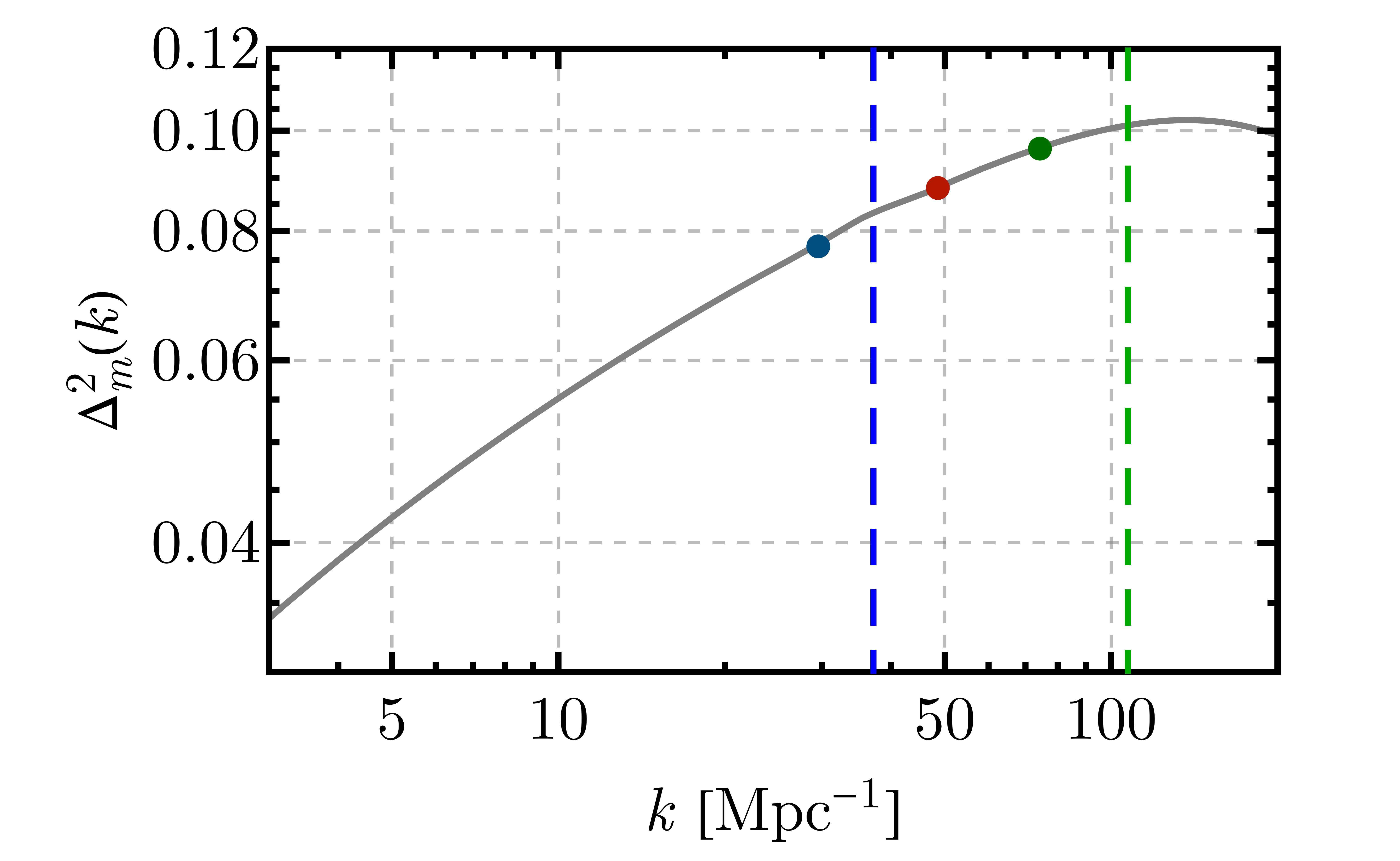}
	\caption{Dimensionless matter power spectrum at $z=20$ (averaged over DM-baryon relative velocities) as a function of wavenumber $k$.
		We highlight three arbitrary bins in colors, with wavenumbers $k=30,50,$ and 75 $\Mpcinv$, which we will use throughout the text to illustrate results.
		The dashed blue and green lines represent the typical wavenumbers of haloes that can form stars at $z=15$ (atomic-cooling) and $z=25$ (molecular-cooling), respectively.
	}	
	\label{fig:Pkbins}
\end{figure}

Throughout this work we will study which set of amplitudes $\mathbf a$ can be measured under different experiments, and how correlated they are to each other.
For illustration purposes, we will often show results for three of our bins, chosen to be centered around $k=\{30,50,75\}$ Mpc$^{-1}$ (each with width $\Delta \log k \approx 0.04$).
These bins exemplify well the overall trends, acting as examples of the effect of a change on low-, medium-, and high-$k$ modes.
We show the fiducial $\Delta^2_{m,\rm fid}(k)$ at $z=20$ in Fig.~\ref{fig:Pkbins}, where we mark with dots the three bins aforementioned.
Additionally, we also mark the typical wavenumber $k_{\rm atom}=37\,\rm Mpc^{-1}$ of atomic-cooling haloes with $M_{\rm atom}=3 \times 10^7 \, M_\odot$ at $z=15$,  
and the maximum wavenumber that we can realistically probe, $k_{\rm mol}=100 \, \rm Mpc^{-1}$ (corresponding to molecular-cooling haloes with $M_{\rm mol}=2 \times 10^6 \, M_\odot$ at $z=25$).
Note that, over this entire range, the variance of matter fluctuations is always below unity at $z=20$, as even small-scale structure has only started forming during cosmic dawn.

\begin{figure}[hbtp!]
	\centering
		\includegraphics[width=0.46\textwidth]{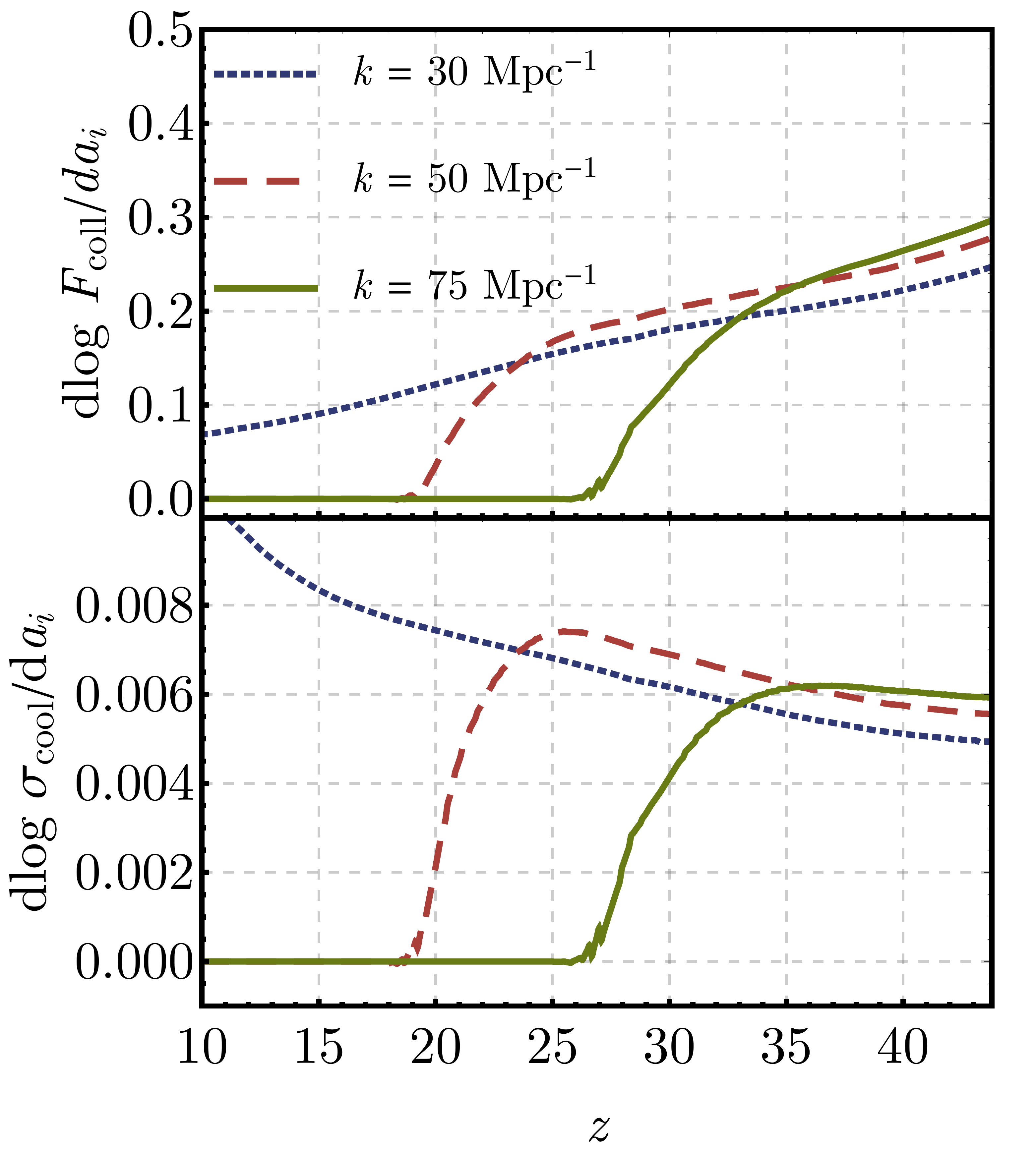}
	\caption{Logarithmic derivative of the collapsed fraction $F_{\rm coll}$ of baryons into star-forming galaxies (\emph{top}) and the (square root of the) variance $\sigma_M$ of star-forming galaxies (\emph{bottom}) as a function of redshift $z$, when changing the matter power spectrum in three different $k$-bins (denoted by their centroid).
	The derivatives die off at the redshift at which the typical haloes sourced from each wavenumber stop forming stars efficiently, and we note that the mass of these haloes spans $M=10^6-10^8 \Msun$ over $z=12-25$.
	}	
	\label{fig:dlogFcoll}
\end{figure}

We show in Fig.~\ref{fig:dlogFcoll} how the collapsed fraction $F_{\rm coll}$ varies as a function of redshift when changing the three bin amplitudes detailed above.
Increasing the power of large-scale modes has a fairly smooth effect across all redshifts, whereas large-$k$ modes preferentially affect $F_{\rm coll}$ at earlier times, and their effects vanish below some redshift ($z\sim20$ for $k=50\Mpcinv$, and $z\sim 25$ for $k=75\Mpcinv$).
That is because as the universe evolves, the small haloes that are formed out of those high-$k$ fluctuations stop being able to form stars (due to feedback), cancelling the effect of these modes on $F_{\rm coll}$.
The situation is similar for $\sigma_{\rm cool}$, defined to be the standard deviation $\sigma_M$ of haloes of mass $M=M_{\rm cool}(z)$, which we also show in Fig.~\ref{fig:dlogFcoll}.

\subsection{The Observable}

So far we have discussed how the small-scale matter power spectrum affects the formation of the first galaxies during cosmic dawn, through quantities such as $F_{\rm coll}$ and $\sigma_{\rm cool}$.
However, these galaxies are challenging to detect with regular surveys, as they are far too distant and dim.
We will, instead, use the 21-cm line of neutral hydrogen as a tracer of star-formation at high redshifts.

Our observable will be the 21-cm brightness temperature, defined as~\cite{Pritchard:2011xb}
\be
T_{21} = 38\,{\rm mK }\, \left(1 - \dfrac{T_{\rm cmb}}{T_s}\right) \left(\dfrac{1+z}{20}\right)^{1/2}\!\!\! x_{\rm HI} (1 + \delta_b) \dfrac{\partial_r v_r}{H(z)}.
\label{eq:T21}
\ee
Here $T_{\rm cmb}$ and $T_s$ are the CMB and spin temperatures, respectively, $x_{\rm HI}$ is the neutral-hydrogen fraction, $\delta_b$ is the baryon overdensity, and the last term includes the gradient $\partial_r v_r$ of the radial velocity and the Hubble parameter $H(z)$.

The basic picture is that the first galaxies impact the state of the IGM around them, allowing us to map them with the 21-cm line of hydrogen.
The first generation of stars filled the universe with UV photons. 
Those with energies between the Lyman-$\alpha$ and Lyman-$\beta$ lines were able to travel significant distances until redshifting into the Lyman-$\alpha$ transition, when they were resonantly scattered by hydrogen.
This scattering couples the microscopic (hyperfine) and macroscopic (thermal) states of hydrogen, equating the spin and gas temperature of hydrogen, in a process called Wouthuysen-Field coupling~\cite{Wout,Field,Hirata:2005mz}.
This makes $T_s$ comparable to the gas temperature $T_g$, which is significantly lower than $T_{\rm cmb}$ in Eq.~\eqref{eq:T21}, and thus gives rise to deep 21-cm absorption of CMB photons.
The first galaxies will also emit abundant X-rays~\cite{Mesinger:2012ys,Fialkov:2014kta}. 
This emission will heat up the IGM around each galaxy, increasing the hydrogen spin temperature, and thus eventually sourcing 21-cm emission, as for $T_s\geq T_{\rm cmb}$ the 21-cm temperature turns positive in Eq.~\eqref{eq:T21}.

It is through these two processes that $T_{21}$ acts as a tracer of the first galaxies across cosmic dawn.
We will obtain all of our theoretical results with the {\tt 21cmvFAST} code~\cite{Munoz:2019rhi}\footnote{\url{https://github.com/JulianBMunoz/21cmvFAST}}, based on {\tt 21cmFAST}~\cite{Mesinger:2010ne,Mesinger:2007pd,Greig:2015qca}\footnote{\url{https://github.com/andreimesinger/21cmFAST}}, but including the effect of DM-baryon relative velocities, which strongly influence the formation of the first stars.
We refer the reader to references  outlined above for details on the specific implementation of the above equations onto this code.

In this work we will use two types of 21-cm observations, with different strengths and weaknesses.
The first are global-signal measurements, which target the overall evolution of the IGM, although they do not have spatial resolution.
The second are measurements of the 21-cm fluctuations, which allow us to more precisely map the distribution of the first stars.
In the next sections we describe each of them.

\section{Global Signal}
\label{sec:global}

We begin by studying the 21-cm global signal, defined to be the average of the 21-cm temperature across the entire sky, as a function of redshift.
Note, in passing, that generally $\VEV{T_{21}} \neq T_{21}(\delta=0)$, from Eq.~\eqref{eq:T21}, as there are significant nonlinearities in $T_s$ as a function of $\delta$.
We overcome this issue by running simulation boxes with {\tt 21cmvFAST}, and averaging over the resulting $T_{21}$ box.
This is in contrast to some previous works on non-CDM models~\cite{Lidz:2018fqo,Schneider:2018xba}, which used the public ARES code~\cite{Mirocha:2014faa} instead of running a simulation and averaging it.

Within {\tt 21cmvFAST}, we use a box that is 900 Mpc in size, with 3 Mpc resolution\footnote{We note that, while the 3-Mpc resolution of the {\tt 21cmvFAST} simulations is rather coarse, the sensitivity to the small-scale ($k\gtrsim 3\,\Mpcinv$) power comes from the sub-grid halo mass functions described in Sec.~\ref{sec:model}. Thus, for the wavenumbers we study here the large-scale ($k\lesssim 3 \Mpcinv$) distribution of matter remains unaltered, whereas the amount of haloes in each simulation pixel changes with the small-scale power.}. 
This is chosen to be large enough to resolve the large-scale 21-cm fluctuations, while keeping enough resolution for the moving-mesh approximations of Ref.~\cite{Tseliakhovich:2010bj} to be valid.
In addition to the power-spectrum bin amplitudes introduced above, the 21-cm signal depends on several astrophysical parameters.
These are the overall fraction $f_*^{(0)}$ of gas that is converted into stars (where we drop the superscript from now on for simplicity), the amplitude $B$ of the feedback strength in Eq.~\eqref{eq:Mcoolfeedback}, the X-ray luminosity per unit of star formation $\log_{10} (L_X)$ (to which we will refer to as $L_X$ unless confusion can arise), and the X-ray threshold $E_0$, below which we assume no X-rays escape to the IGM.
These four parameters have to be fit simultaneously to the matter power spectrum, and thus we will marginalize over them in the rest of this work.
This is an improvement over previous work, which either fixed astrophysical parameters or varied only the gas fraction.
There is an additional X-ray parameter, the spectral index $\alpha_X$, which produces only a modest change in $T_{21}$~\cite{Greig:2017jdj}, and thus we do not vary it here.
We have chosen to work with the astrophysical parametrization of {\tt 21cmFAST} for convenience, but we note that other parametrizations are possible~\cite{Mirocha:2015jra,Park:2018ljd}.

\begin{figure}[hbtp!]
	\centering
	\includegraphics[width=0.46\textwidth]{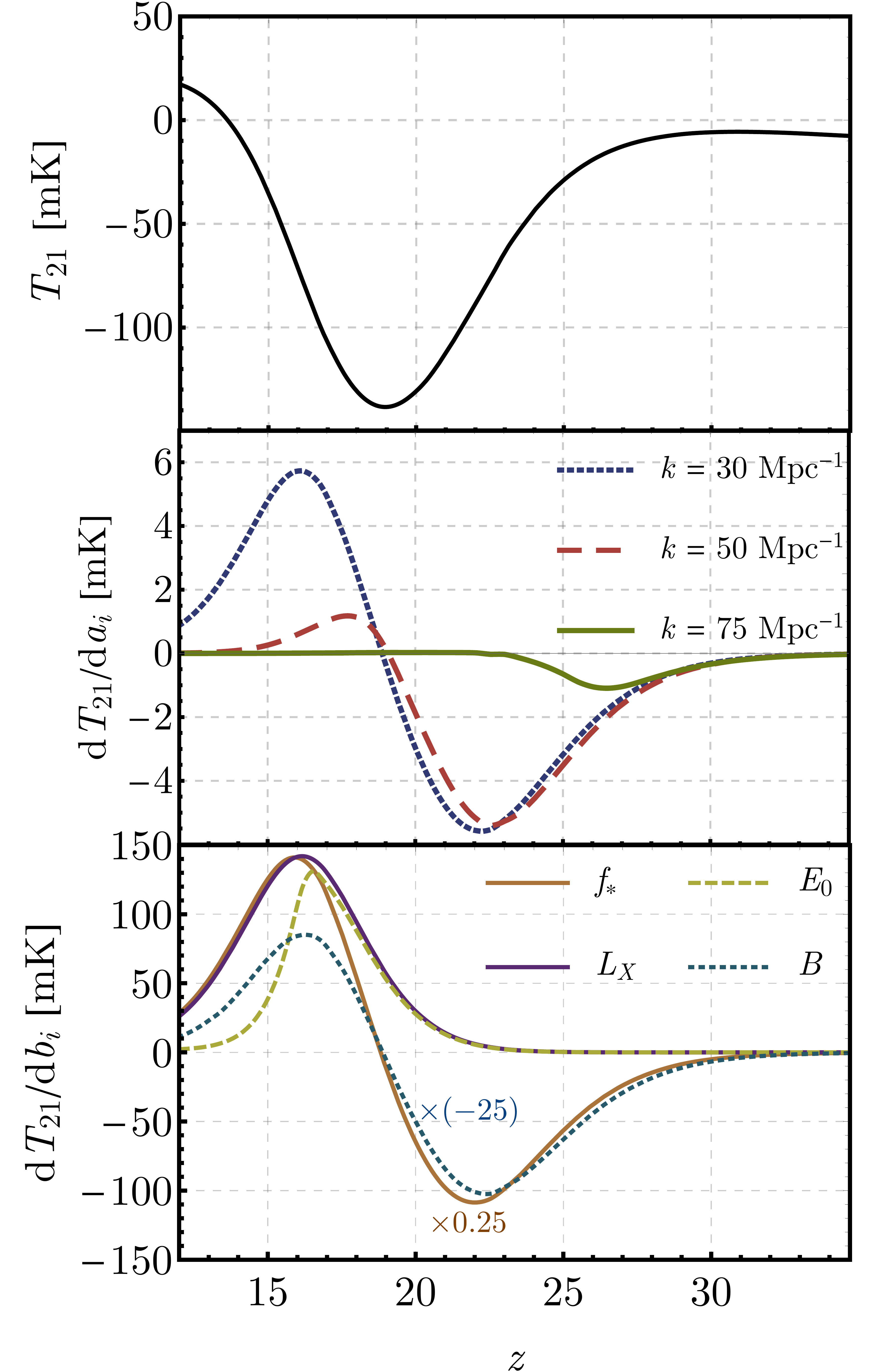}
	\caption{{\bf Top}: Fiducial 21-cm global signal as a function of redshift $z$ across cosmic dawn, obtained with {\tt 21cmvFAST}.
	The first stars formed at $z\approx 25$, which produced 21-cm absorption, turning into emission at $z\approx 14$. \newline
	{\bf  Middle}: Derivative of the global signal with respect to three bin amplitudes as a function of redshift.
	As in Fig.~\ref{fig:dlogFcoll}, high-$k$ bins only affect high-$z$ signals. \newline
	{\bf Bottom}: Derivative of the global signal with respect to our astrophysical parameters.
	In this plot, and for illustration purposes only, we have rescaled the $f_*$ (stellar fraction) and $B$ (feedback amplitude) derivatives by factors of 0.25 and $-25$, respectively, to show them in the same scale.
	}	
	\label{fig:T21global}
\end{figure}

We show the 21-cm global signal in our fiducial case in Fig.~\ref{fig:T21global}, where we can distinguish the two distinct sub-eras that comprise cosmic dawn.

First, the Lyman-$\alpha$ coupling era (LCE), lasting from $z=27-19$, from the first stellar formation until the minimum of $T_{21}$, when X-ray heating starts to dominate.
During this era the first stars emit UV photons which, after redshifting into the Lyman-$\alpha$ transition, produce Wouthuysen-Field coupling between the hydrogen spin and kinetic temperature, giving rise to (anisotropic) 21-cm absorption.

Second, the epoch of heating (EoH), which starts at $z\simeq19$ and does not end until $z\simeq14$, at which point the 21-cm signal crosses zero. In this epoch the X-rays produced in the first galaxies heat up the hydrogen, inhomogeneously reducing the amount of 21-cm absorption, until it turns into emission.
After the gas is fully heated, the hydrogen will start getting reionized by the UV emission from subsequent generations of stars.
This marks the beginning of the epoch of reionization (EoR), during which the global signal smoothly decays to zero as the neutral-hydrogen fraction slowly vanishes~\cite{Furlanetto:2004nh,McQuinn:2005hk}.
We do not study the EoR, as more-complex feedback processes can drive the minimum mass of star-forming haloes to larger values~\cite{MoGalaxyFormation,loeb2013first}.

We will now find how the 21-cm global signal changes when different parameters are varied.
In principle we should perform a Markov-chain Monte Carlo (MCMC) search in parameter space, running {\tt 21cmvFAST} for each combination of parameters.
That is, however, unfeasibly slow, as the simulations take significant time ($\sim$ hrs in a single core). 
While this hurdle can potentially be circumvented with techniques such as emulation~\cite{Kern:2017ccn}, here instead we choose the simpler approach of varying the parameters around our fiducial model, in order to compute a Fisher matrix.
In addition to being computationally simple, this has the advantage of preserving all the input modelling (whereas emulators or other interpolators can be imprecise).
We calculate the numerical derivative for each of our parameters by using the {\tt 21cmvFAST} code, and the $F_{\rm coll}$ and $\sigma_{\rm cool}$ computed in the previous section.
In the following subsections we outline the derivative of $T_{21}$ with respect to each of the relevant parameters.

\subsection{Small-Scale Matter Power Spectrum}

We begin by calculating the derivative of $T_{21}$ with respect to the bin amplitudes, which we show in Fig.~\ref{fig:T21global} for the same three bins that we are using as examples throughout (the ones marked in Fig.~\ref{fig:Pkbins}).
Broadly speaking, increasing the power in any of these bins produces more stars, and thus deeper absorption during the LCE ($z>19$)---which yields a negative derivative.
This behavior is reversed during the EoH, where more stars produce more X-rays, and thus a positive derivative.

Note that, as was the case for Fig.~\ref{fig:dlogFcoll}, while the low-$k$ bin affects the 21-cm signal throughout all cosmic dawn (roughly as an overall normalization), the largest-$k$ bins only alter the signal at high redshifts, when small haloes were forming stars.
This will allow us to distinguish different $k$-bins, and thus measure the matter power spectrum at different scales.

\subsection{Astrophysical Parameters}

In addition to the cosmological parameters of interest (the bin amplitudes $\mathbf a$), we marginalize over the known astrophysical variables outlined above.
These can be divided in two categories.

First, the fraction $f_*$ of collapsed baryons that convert into stars and the feedback amplitude $B$.
These two parameters, with fiducial values of 0.1 and 7 respectively, act as (redshift- and mass-dependent) normalizations of the collapsed fraction, and are meant to represent our ignorance about the first stellar formation, including the effect of Lyman-Werner feedback.

Second, the X-ray spectra of the first galaxies is poorly constrained, and both the X-ray luminosity $L_X$ per unit of star-formation, and the X-ray cutoff energy $E_0$, below which we assume no emission, have to be marginalized over.
These two parameters, chosen to have fiducial values of 40 and 0.2 keV, respectively, only change the signal during the EoH.

We show the derivatives of $T_{21}$ with respect to these four parameters in Fig.~\ref{fig:T21global}, as a function of redshift.
The derivative for $f_*$ is very similar to that of the lowest $k$-bin, as both act as an overall normalization.
For the feedback amplitude $B$, however, the effect is more marked at high redshifts, where Lyman-Werner feedback has a stronger effect on the amount of stars formed.
Note that in Fig.~\ref{fig:T21global} we have changed the sign of the $B$ derivative for easier comparison with $f_*$, as more feedback produces fewer stars (and thus it has the opposite sign of $f_*$).
The X-ray parameters ($L_X$ and $E_0$), on the other hand, only affect the 21-cm signal when X-rays are important ($z\lesssim 20$), as expected, speeding up the heating of the IGM, and thus showing a positive derivative.
Note that, while $L_X$ and $E_0$ have a similar effect during the onset of the EoH ($z=17-20$), the effect of $E_0$ decreases in the later stages ($z\lesssim 17$), which will allow us to disentangle these two parameters.

\subsection{Foregrounds}

Any global-signal experiment has to unearth the cosmological 21-cm signal from the much-larger radio foregrounds present in the data.
Given arbitrary freedom, said foregrounds would be able to fit any set of data, and thus it would be impossible to detect the primordial signal~\cite{Liu:2012xy}.
Nonetheless, it is expected that the foreground emission is a smooth function of frequency, and thus its behavior can be encapsulated into a few terms.
Here we will specify these terms, in order to marginalize over their amplitudes

\begin{figure}[hbtp!]
	\centering
	\includegraphics[width=0.46\textwidth]{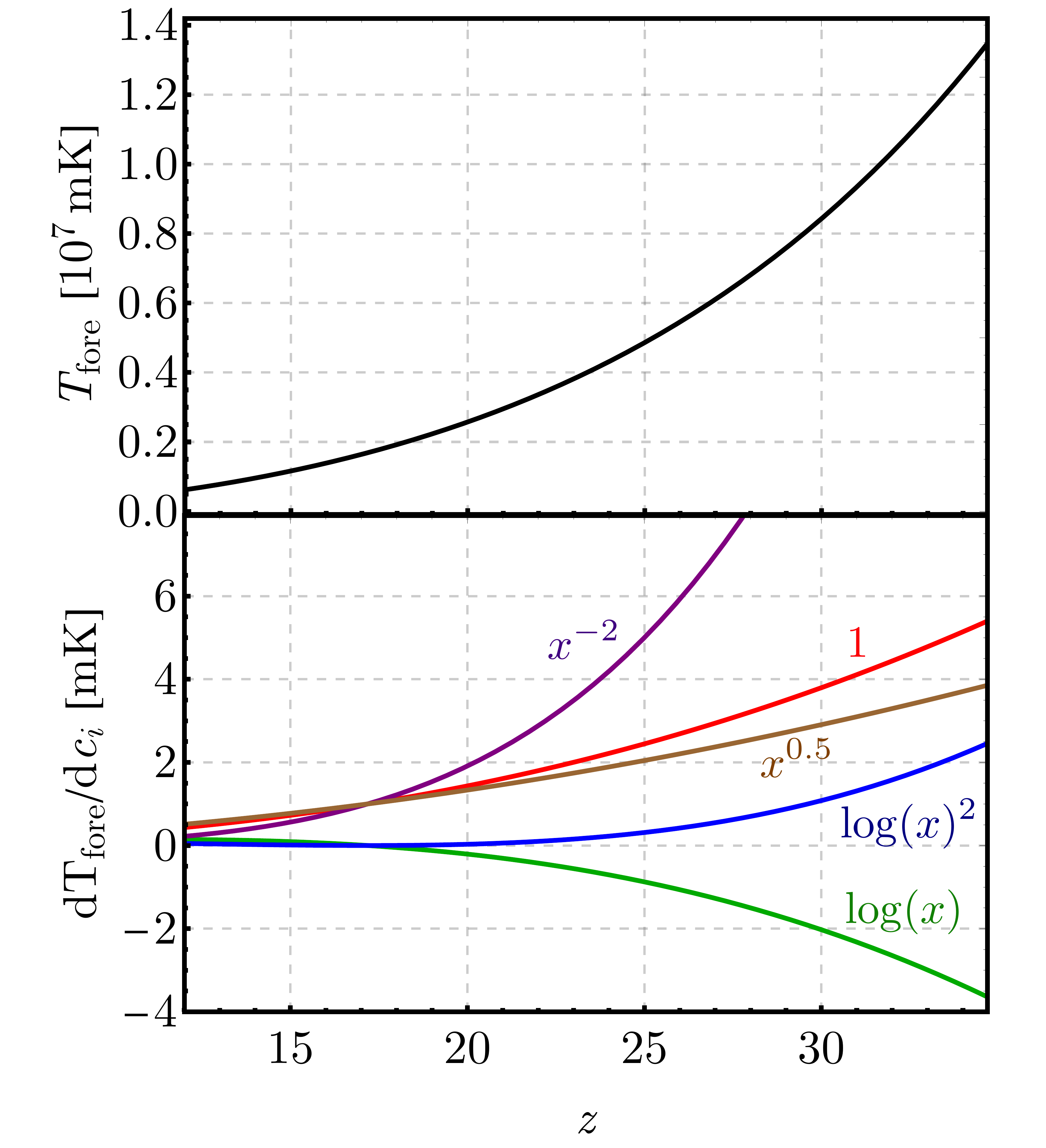}
	\caption{{\bf Top}: Foreground 21-cm temperature as a function of redshift, from Eq.~\eqref{eq:fore}, where we have translated frequency into redshift.
	{\bf  Bottom}: Derivative of the 21-cm foreground temperature with respect to the amplitudes $c_i$ of each component $g_i(x)$, which we will marginalize over. As in the main text, $x\equiv \nu(z)/\nu_0$ with $\nu_0=72$ MHz.
	}	
	\label{fig:T21fore}
\end{figure}

The exact form of the foreground signal $T_{21}^{\rm fore}$ during cosmic dawn has not been fully established.
Here we follow the five-term model used in Ref.~\cite{Bowman:2018yin}, where
\be
T_{21}^{\rm fore}(\nu) = \left (\dfrac {\nu}{\nu_0} \right)^{-2.5}\!\!\! \times \sum_{i=1}^5 c_i g_i(\nu/\nu_0),
\label{eq:fore}
\ee
where $g_i(x)=\{1,\log(x),\log(x)^2,x^{0.5},x^{-2}\}$ is our basis of functions, and $c_i$ are coefficients that we will marginalize over.
We take the fiducial values of $\pmb c = \{{1570, 700, -1200, 750, -175}\}$ K, from Ref.~\cite{Bowman:2018yin}, and show the resulting $T_{21}^{\rm fore}(\nu)$ in Fig.~\ref{fig:T21fore}.
Note that Eq.~\eqref{eq:fore} is a linearization of a model that includes both atmospheric and Galactic foreground effects, around some central frequency, chosen to be $\nu_0 = 72$ MHz. 
This suffices for our purposes, as we will take derivatives in order to compute the Fisher matrix.
We show said derivatives ($dT_{\rm fore}/dc_i$) in Fig.~\ref{fig:T21fore}, where it is clear that they are all smooth functions, although they will still be correlated with the different parameters introduced above, and thus we have to marginalize over those foregrounds at the same time as the primordial signal.

\subsection{Results}

In order to find out how well each of the parameters introduced above can be measured, we need to consider experimental errors.
We will assume an EDGES-\emph{like} experiment, with a frequency resolution of $\Delta \nu = 0.4$ MHz and $1000$ hours of observation over the frequency range $\nu=50-110$ MHz (corresponding to $z\approx12-27$).
Then, we can find the uncertainty on each frequency bin $i$ using the radiometer equation~\cite{Pritchard:2010pa},
\be
\sigma_i = \dfrac{T_{\rm sky} (\nu_i)}{\sqrt{\Delta\nu t_{\rm obs}}},
\ee
where $t_{\rm obs}$ is the observation time, and
\be
T_{\rm sky} (\nu) = T_{21} (\nu) + T_{\rm fore}(\nu)  
\ee
is the total sky temperature, to which we could add any potential systematic effects inherent to each instrument.

Using the specifications outlined above, we can find the signal-to-noise ratio of our fiducial 21-cm signal (from Fig.~\ref{fig:T21global}) to be
\be
{\rm SNR} = \left[\sum_i \dfrac{T_{21}^2(\nu_i)}{\sigma_i^2} \right]^{1/2} \approx 560.
\ee
This number, however, does not determine the significance at which a signal can be extracted from the foregrounds, as those have to be fitted for at the same time.
Moreover, in order to forecast errors for each parameter we have to account for correlations between them as well.

Thus, we construct a Fisher matrix by summing over frequencies as~\cite{Pritchard:2010pa,Liu:2015gaa}
\be
F_{\alpha \beta} = \sum_i \dfrac{\partial T_{\rm sky}(\nu_i)}{\partial \theta_\alpha} \dfrac{\partial T_{\rm sky}(\nu_i)}{\partial \theta_\beta} \sigma_i^{-2},
\ee
where the parameter list $\pmb \theta$ includes the bin amplitudes ($\pmb a$), astrophysical parameters  ($\pmb b$), and foregrounds ($\pmb c$), i.e.,  $\pmb \theta = \{\mathbf a, \mathbf b, \mathbf c\}$.
Given the Fisher matrix for our experiment, we can estimate the marginalized error bars for each parameter by inverting it, so that $\sigma^2(\theta_\alpha) = (F^{-1})_{\alpha \alpha}$.

\subsubsection{Astrophysical Parameters}

\begin{figure}[t!]
	\centering
	\includegraphics[width=0.52\textwidth]{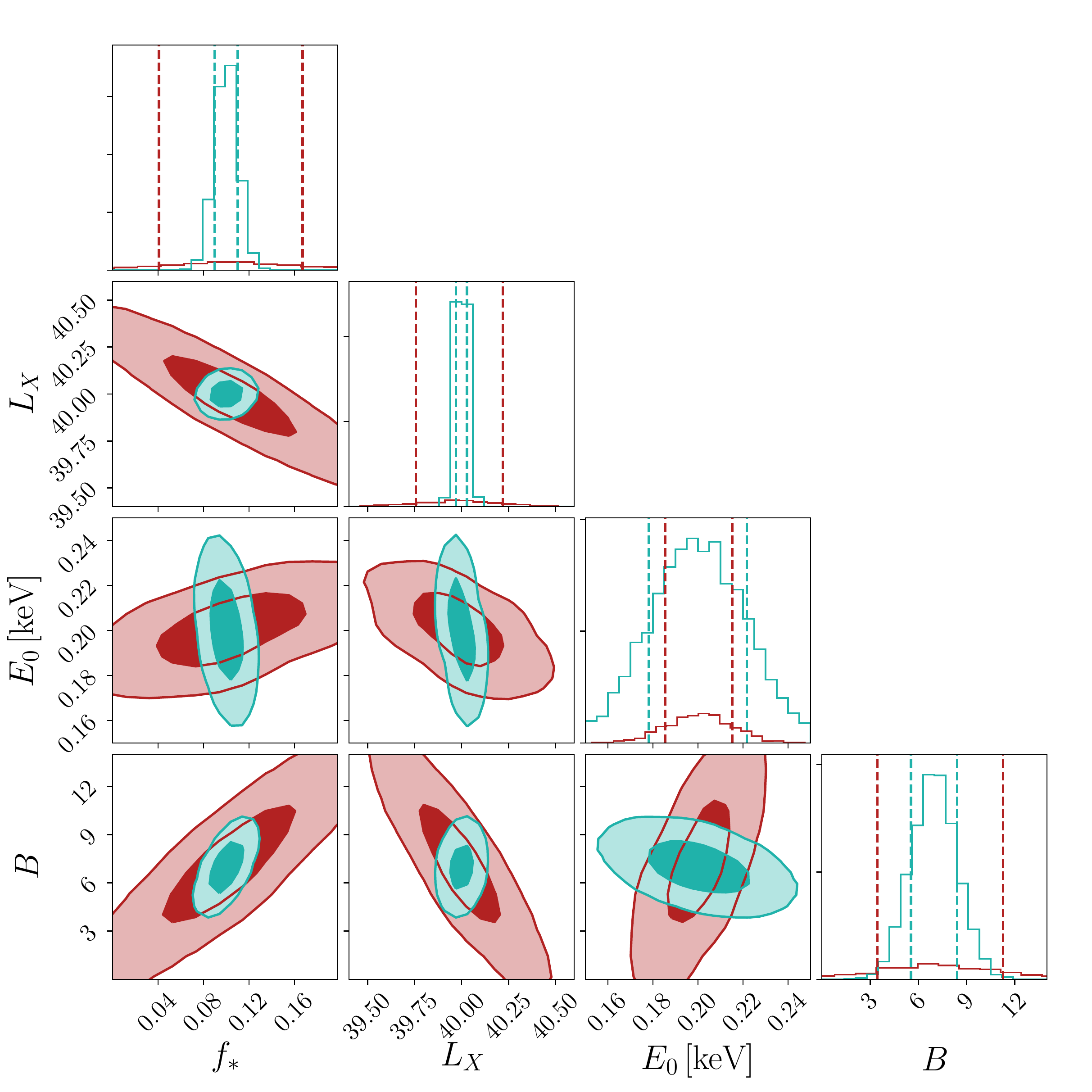}
	\caption{Forecasted confidence ellipses (68\% in dark and 95\% in light color) for the four astrophysical parameters, given an EDGES-like global-signal experiment (red), and a HERA-like fluctuation experiment under moderate foregrounds (teal).
	While 21-cm fluctuations detect the cosmic signal at a lower signal-to-noise ratio, they can better break degeneracies between parameters, yielding better constraints.
	}
	\label{fig:Astro_ellipses}
\end{figure}

We start by fixing all $a_i=0$ (i.e., assuming CDM), and finding the errors in the astrophysical and foreground parameters.
In this case, an EDGES-like experiment, as the one we are employing, could measure all the astrophysical parameters with $\mathcal O(1)$ relative precision, when marginalizing over the foreground amplitudes.
For instance, we find $\sigma(f_*) = 0.06$ and $\sigma(B) =4$, which are large uncertainties compared with the fiducial values of 0.1 and 7 for those two parameters.
That is because of the degeneracy between them, as they both control the amount of star formation.
The X-ray parameters, on the other hand, can be measured at greater precision, as $\sigma(\log_{10}(L_X)) = 0.3$ and  $\sigma(E_0) =0.02$ keV, to be compared with their values of 40.0 and 0.2 keV, respectively. 
We show the 2-dimensional probability distribution functions of the astrophysical parameters in Fig.~\ref{fig:Astro_ellipses}, obtained using the {\tt emcee} and {\tt corner} packages~\cite{emcee, corner}.
Our results show that a global-signal experiment can only measure the astrophysical parameters to $\mathcal O(1)$ uncertainty, given the large degeneracies between them.
This is comparable to the MCMC-based results of Ref.~\cite{Monsalve:2018fno}, using the EDGES high-band data.

\subsubsection{Matter Power Spectrum}

We now vary over the bin amplitudes as well.
Given the large degeneracies between contiguous bins, we will join them in larger bins, occupying a wider $k$ range.
We divide our range in three large bins, spanning $k=(20-38)$ Mpc$^{-1}$, $k=(38-80)$ Mpc$^{-1}$, and $k=(80-108)$ Mpc$^{-1}$.
These ranges are designed to be fairly uncorrelated with each other, and are inspired by the principal-component analysis that we will detail later.
In this case, our cosmological parameter list is simply $\mathbf{ \tilde a_i}=\{\tilde a_1,\tilde a_2,\tilde a_3\}$, where $\tilde a_i$ is the amplitude of each of the broad bins above, with fiducial value of unity in CDM.

These $\tilde a_i$ parameters can be highly correlated with the astrophysical parameters, so in order to keep the value of the astrophysical parameters strictly positive, we now include a broad prior of $f_*=0.1\pm0.1$, $B=7\pm7$, and $E_0=(0.2\pm0.2)$ keV (and no prior on $L_X$).
Without these priors, unphysical values of these parameters could be reached, and the errors on the bin amplitudes would be significantly larger, given the degeneracies.
Nonetheless, with the physicality prior imposed above, it will be possible to measure the matter power spectrum in the lowest two $k$-bins, where we find $\sigma(\tilde a_1) = 0.82$ and $\sigma(\tilde a_2) = 0.56$.
The last (highest $k$) bin, however, will not be measurable at any precision, as $\sigma(\tilde a_3) = 45\gg 1$.

\begin{figure}[btp!]
	\centering
	\includegraphics[width=0.46\textwidth]{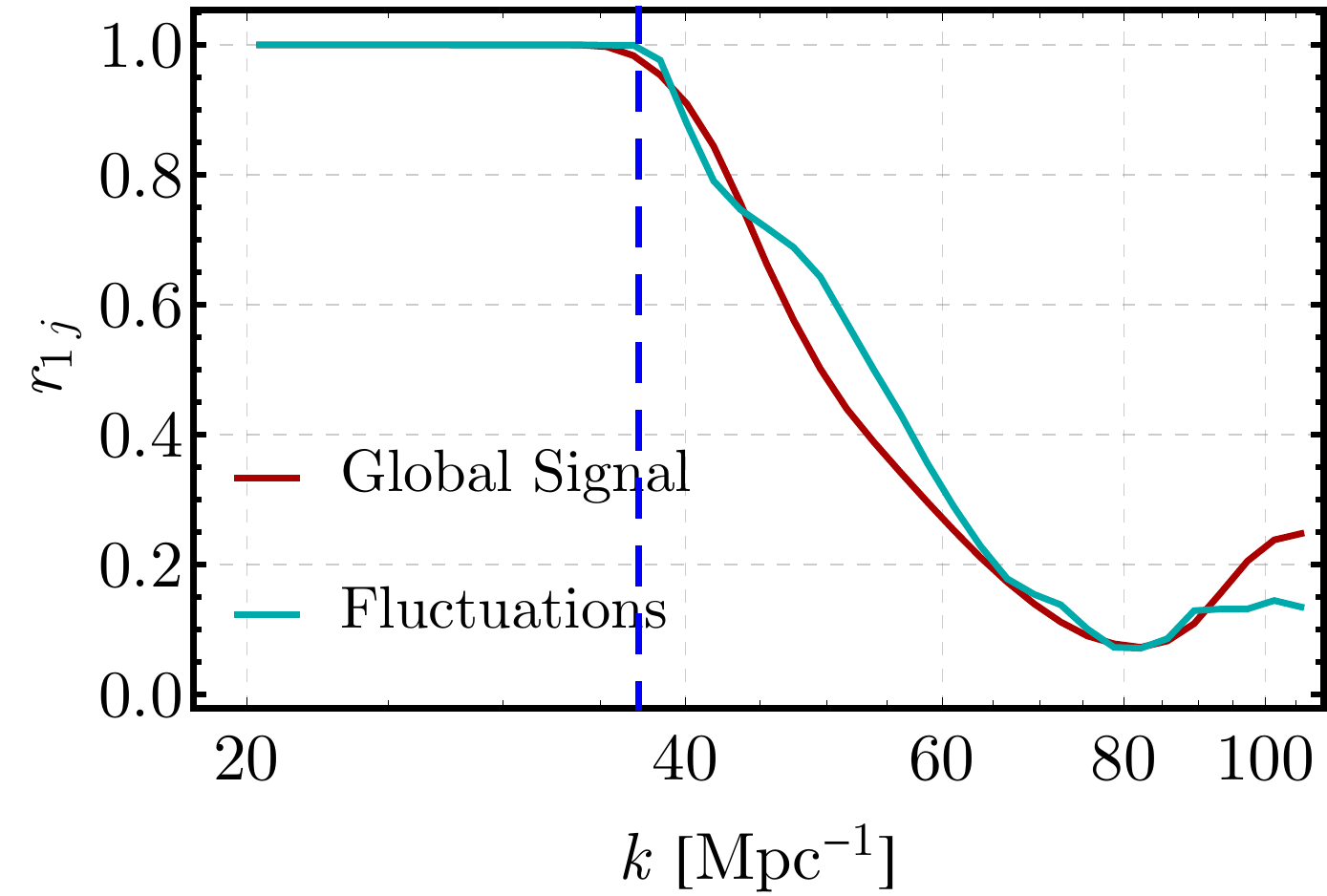}
	\caption{Fisher correlation parameter $r_{1j}$, defined as in Eq.~\eqref{eq:rfisher}, between our first wavenumber (at $k=20\Mpcinv$) and the rest.
	All modes with $k<k_{\rm atom}$ (denoted by a vertical blue dashed line) are highly correlated, as they act as a normalization for both 21-cm signals.
	}	
	\label{fig:Corrfirstbin}
\end{figure}

In our simulations we have only varied the amplitude of matter wavenumbers down to $k_{\rm min}=20$ Mpc$^{-1}$.
Nonetheless, all modes below $k_{\rm atom} \sim 40$ Mpc$^{-1}$ have the same effect on 21-cm observables.
That is because these modes can all affect the abundance of atomic-cooling haloes, which by the end of our simulations (at $z=12$) are still able to form stars, and thus all the modes change the 21-cm signal in a very similar fashion.
To illustrate this point, we show in Fig.~\ref{fig:Corrfirstbin} the Fisher correlation parameter, defined as
\be
r_{ij} = \dfrac{F_{ij}}{\sqrt{F_{ii}{F_{jj}}}},
\label{eq:rfisher}
\ee
between our first narrow bin (spanning $k=(20-20.9)$ Mpc$^{-1}$) and the rest of them, as a function of the narrow-bin centroid $k$.
This correlation is nearly unity for $k<k_{\rm atom}$, and quickly decreases afterwards (and for comparison, the correlation between this first bin and $f_*$ is nearly unity both for the 21-cm global signal, $r=0.994$, and the 21-cm fluctuations described below, $r=0.943$).
Then, the 21-cm signal will be sensitive to all modes below $k_{\rm atom}$ equally, which allows us to trivially extend our analysis to the last currently known measurement of the matter power spectrum (from the Lyman-$\alpha$ forest~\cite{Chabanier:2019eai}), which is at $k_{\rm Ly-\alpha}\approx 3$ Mpc$^{-1}$.
We have checked that changing the value of $k_{\rm min}$ only alters the precision at which the amplitude $\tilde a_1$ of the first bin can be measured, scaling as $\log(k_{\rm atom}/k_{\rm min})$, and not its correlation with any other parameter.
Therefore, we take the first bin to extend over the range $k=(3-38)\Mpcinv$, and thus forecast an uncertainty in its amplitude of $\sigma(\tilde a_1) = 0.2$.

\begin{figure}[hbtp!]
	\centering
	\includegraphics[width=0.46\textwidth]{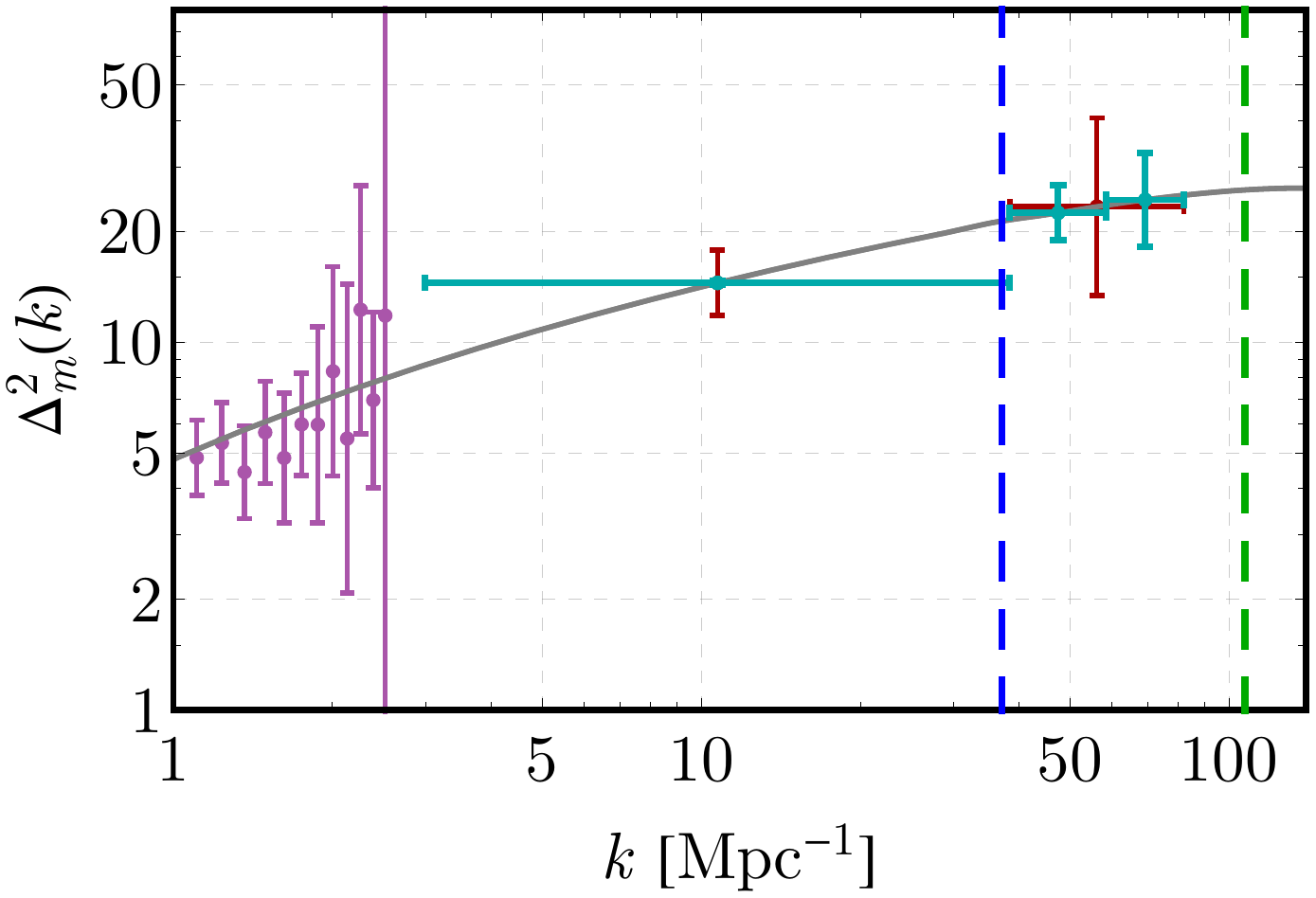}
	\caption{Dimensionless matter power spectrum as a function of wavenumber $k$, linearly extrapolated to $z=0$.
	The purple points represent Lyman-$\alpha$ measurements from Ref.~\cite{Chabanier:2019eai}. 
	The red crosses show our forecasted constraints for a global-signal EDGES-like 21-cm experiment, marginalizing over the other bin amplitudes, as well as astrophysical and foreground parameters.
	The teal crosses show the forecasted error bars  assuming a HERA-like 21-cm fluctuation experiment, also marginalizing over astrophysical parameters, and assuming moderate foregrounds.
	The lowest $k$-bin is the same in both cases (covering $k=(3-38) \Mpcinv$), whereas the two higher bins in the power-spectrum case (from $k=(38-60) \Mpcinv$ and $k=(60-80) \Mpcinv$) are joined into one when considering the global signal.
	Note that the forecasted error bar for the lowest-$k$ using 21-cm fluctuations is 1.2\%, and thus appears smaller than the line width.
	As before, vertical blue and green dashed lines show the typical wavenumbers of atomic- and molecular-cooling haloes.
	}	
	\label{fig:Pmz0_bins}
\end{figure}

We illustrate our results by showing the matter power spectrum, linearly extrapolated to $z=0$, in Fig.~\ref{fig:Pmz0_bins}.
Our forecasted error bars reach larger $k$ values than the strongest current Lyman-$\alpha$ constraints, from Ref.~\cite{Chabanier:2019eai}.
Note, however, that the lowest $k$-bin covers a wide range of wavenumbers, so coherent changes in the power spectrum (e.g., oscillations) would evade detection.
The bin centered at $k\sim 50\Mpcinv$, on the other hand, can be measured within a narrower band, allowing for precise constraints on DM models, as we will explore below.

\subsection{EDGES data}

Before leaving the global-signal studies, let us find what constraints can be achieved with the first claimed 21-cm detection, by the EDGES collaboration~\cite{Bowman:2018yin} (see, however, Refs.~\cite{Hills:2018vyr,Bradley:2018eev} for criticisms regarding foregrounds and systematics).
This first claimed detection is centered around $\nu=78$ MHz, with a width of $\Delta \nu = 20$ MHz (corresponding to $z\approx 15-20$).
While the reported amplitude $A\sim -500$ mK of the signal is anomalous (as it is twice as deep as allowed, see Refs.~\cite{Munoz:2018pzp,Barkana:2018lgd,Pospelov:2018kdh,Ewall-Wice:2018bzf}), this detection is only possible if enough stars exist by $z=20$, as they are required to emit the necessary Lyman-$\alpha$ photons for Wouthuysen-Field coupling.

Here, as in Refs.~\cite{Lidz:2018fqo,Schneider:2018xba,Safarzadeh:2018hhg,Leo:2019gwh,Boyarsky:2019fgp,Lopez-Honorez:2018ipk,Yoshiura:2018zts}, we take advantage of this lower bound on the amount of Lyman-$\alpha$ photons to probe different dark-matter models.
We will phrase the EDGES requirement in terms of the dimensionless coupling coefficient $x_\alpha$, which enters in the definition of the spin temperature (ignoring collisional coupling during cosmic dawn) as
\be
T_S^{-1} = \dfrac{T_\gamma^{-1} + x_\alpha T_\alpha^{-1}}{1+x_\alpha},
\ee
where $T_\gamma$ is the CMB temperature, and $T_\alpha$ is the color temperature, which is related (and similar in value) to the gas temperature~\cite{Hirata:2005mz}.
By detecting 21-cm absorption, the EDGES result demands $x_\alpha\gtrsim1$ at $z=20$, where the absorption signal starts.
We note that, in addition, the non-detection of signal for $z>20$ implies an upper bound on $x_\alpha$, and thus on the power spectrum~\cite{Yoshiura:2019zxq}, although we will not attempt to find this value here.
Moreover, there is further information contained on the shape of the EDGES signal (for instance the flattened bottom), which however we will not attempt to fit here.

\begin{figure}[hbtp!]
	\centering
	\includegraphics[width=0.46\textwidth]{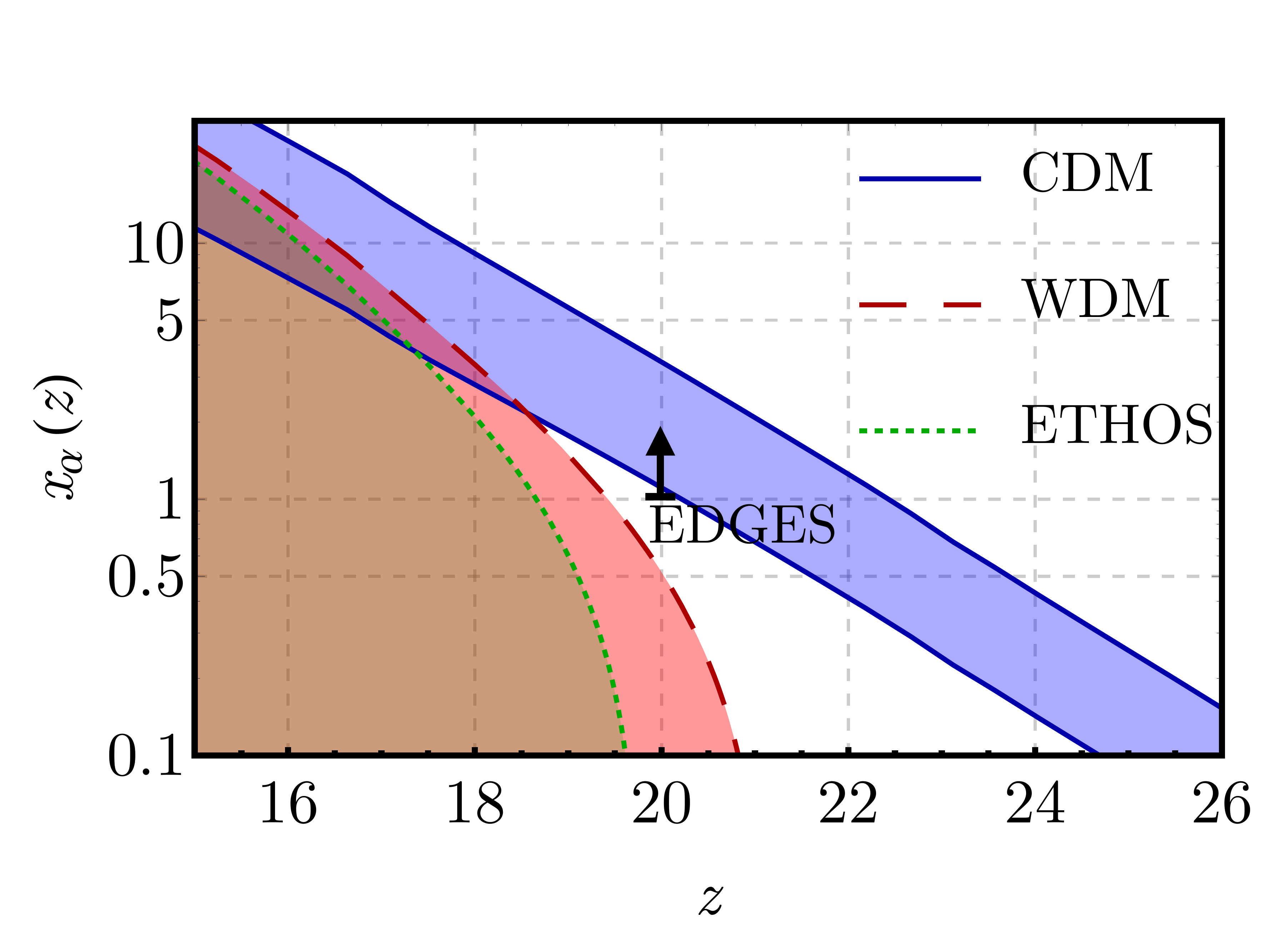}
	\caption{Dimensionless Lyman-$\alpha$ coupling coefficient as a function of redshift.
	The colored bands represent the predictions for different DM models when varying $f_*$ by 50\% around its fiducial value.
	Blue is CDM, red dashed line is WDM with a 5-keV mass, and green dotted line is a model with DM-DR Yukawa-like interactions, obtained by setting $\mathfrak{a}_4=40\Mpcinv$ in ETHOS, a parameter that determines the size of the DM-DR interactions.
	The EDGES measurement requires $x_\alpha$ to be above unity by $z=20$, which is denoted by the black data point, and thus both nCDM models shown are ruled out. 
	Note that this argument only relies on the timing of the EDGES signal, and not on its depth, which is anomalously large.
	}	
	\label{fig:xaEDGES}
\end{figure}

We obtain the global (sky-averaged) value of $x_\alpha$ from our {\tt 21cmvFAST} simulations, and plot it as a function of redshift in Fig.~\ref{fig:xaEDGES}.
Computationally, we obtain $d x_\alpha/da_i$ for each of the matter power spectrum bin amplitudes $a_i$ from our simulations, and construct the $x_\alpha$ for any non-CDM model as 
\be
x_{\alpha,\rm nCDM} = x_\alpha^{\rm CDM} + \sum_i [T_{\rm nCDM}^2(k_i) -1]\dfrac{d x_\alpha}{da_i},
\ee
where $T_{\rm nCDM}=\Delta_{m,\rm nCDM}/\Delta_m$ is the ``transfer function" of the non-CDM model, which accounts for the deviation from CDM.

We will illustrate the constraining power of EDGES data with two simple nCDM examples. 
First, we will assume a warm dark matter (WDM) candidate, in which case the small-scale power is erased by the free streaming of DM particles.
We use the fit to $T_{\rm nCDM}$ from Ref.~\cite{Bode:2000gq},
\be
T_{\rm WDM} (k) = (1 + (\alpha k)^{2 \nu})^{-5/\nu},
\ee
with $\nu=1.2$, and the suppression scale is given by~\cite{Viel:2005qj}
\be
\alpha = \dfrac{0.05}{h} \left( \dfrac{\Omega_{\rm WDM}}{0.25}\right)^{0.11} 
\left( \dfrac{h}{0.7}\right)^{1.22} 
\left( \dfrac{m_{\rm WDM}}{1\,\rm keV}\right)^{-1.11}\rm Mpc.
\ee
We will vary $m_{\rm WDM}$ until it is ruled out by EDGES.

Second, we will consider interactions between DM and dark radiation (DR), which give rise to suppression of fluctuations and dark acoustic oscillations (DAOs)~\cite{CyrRacine:2012fz,Cyr-Racine:2013fsa}---akin to Silk damping and baryon acoustic oscillations (BAOs) in the visible sector.
We will work within the effective theory of structure formation (ETHOS) framework~\cite{Cyr-Racine:2015ihg,Vogelsberger:2015gpr} and focus on the $n=4$ case, which corresponds to a Yukawa-like interaction with a massive mediator. Such a model has been argued to solve some of the small-scale CDM puzzles (see e.g.~Refs.~\cite{Aarssen:2012fx,Chu:2014lja,Kaplinghat:2015aga,Huo:2017vef}). In this case, the opacity between DR and DM takes the form
\begin{equation}\label{eq:ETHOS_4}
    \dot{\kappa}_{\rm DR-DM} = -(\Omega_{\rm DM} h^2)  \mathfrak{a}_4 \left(\frac{1+z}{1+z_{\rm D}}\right)^4,
\end{equation}
where the parameter $\mathfrak{a}_4$ determines the strength of the DM-DR interaction (normalized at the conventional redshift $z_{\rm D}=10^7$). We will increase the interaction strength (through $\mathfrak{a}_4$) until the model is inconsistent with EDGES.
We set the DR temperature to half that of the CMB, which is a natural value if the two sectors were coupled above the weak scale \cite{Fan:2013yva},
and assume that all the DM, with mass $m_\chi=100$ GeV, interacts with it.
For this case, we use {\tt ETHOS-camb}\footnote{\url{https://bitbucket.org/franyancr/ethos_camb}}~\cite{Cyr-Racine:2015ihg} and numerically fit for the transfer function ratio to CDM.
Note that we do not include the effect of DM self interactions, which can be present in some of these models~\cite{Tulin:2017ara}, but instead only account for the suppression in the matter power spectrum due to interactions with DR \cite{Boehm:2001hm,Boehm:2004th,Krall:2017xcw}.

The properties of the first stars are not currently fully known~\cite{Schauer:2019ihk}, and have to be fit jointly to any deviation from CDM.
We account for this by varying the stellar fraction $f_*$ by 50$\%$ around its fiducial value (of 0.1), as shown in Fig.~\ref{fig:xaEDGES}.
The result is that, while CDM is broadly consistent with EDGES regardless of $f_*$ (as $x_\alpha$ is above unity at $z=20$), WDM with $m_{\rm WDM}\leq5$ keV is not, as even with the largest $f_*$ allowed, $x_\alpha$ is below unity at $z=20$.
This constraint is similar to that obtained in Ref.~\cite{Schneider:2018xba} (where $m_{\rm WDM}\leq 6$ keV).
For the ETHOS case, we find that models with interactions stronger than $\mathfrak{a}_4=40\Mpcinv$ are ruled out, where, for reference, values as large as $\mathfrak{a}_4\approx 300\Mpcinv$ are required to solve the small-scale problems~\cite{Vogelsberger:2015gpr}.

Note that our constraints would be weaker if $f_*$ was allowed to be higher.
Nonetheless, while $f_*$ can fluctuate by orders of magnitude from one galaxy to another, reaching values near unity~\cite{Boyarsky:2019fgp}, the observed $f_*$ is the average over all star-forming galaxies, and thus should not be close to unity.
Additionally, we caution the reader that our ETHOS constraint is only approximate, as the sharp-$k$ filter employed here can fail to reproduce the halo mass function form simulations in cases with DAOs~\cite{Sameie:2018juk}.

 \section{21-cm Fluctuations}
 \label{sec:fluctuations}
 
We have thus far analyzed the 21-cm global signal, corresponding to the average 21-cm absorption or emission over the entire sky, as a function of redshift.
The 21-cm signal has large spatial fluctuations, which can be detected by interferometers and provide angular information in addition to the redshift behavior of the signal.
Let us now study what can be learned about the matter power spectrum from these 21-cm fluctuations.

Detecting 21-cm fluctuations requires a larger experimental effort than the global signal, as hundreds of antennae are needed.
Nonetheless, this observable has two important advantages.
First, as we will show below, the 3D spatial information allows us to better break degeneracies between parameters, yielding better measurements.
Second, while the 21-cm global signal is a relatively low-cost observable (requiring only one antenna and a few hundred hours of observation), its interpretation is highly challenging, as the signal is swamped by foregrounds.
The 21-cm fluctuations do not suffer from that problem, as the foregrounds are expected to be limited to a ``wedge"~\cite{Liu:2009qga,Morales:2012kf,Datta:2010pk,Parsons:2012qh} in $k$ space.

In this work we will focus on upcoming measurements of the 21-cm power spectrum from the hydrogen epoch of reionization array (HERA)~\cite{DeBoer:2016tnn}, currently under construction in South Africa.
A similar analysis could be carried out for other interferometers, such as the low-frequency array (LOFAR)~\cite{vanHaarlem:2013dsa}, the long-wavelength array (LWA)~\cite{Eastwood:2019rwh},
the Murchison widefield array (MWA)~\cite{Beardsley:2016njr},
or the square-kilometer array (SKA)~\cite{Koopmans:2015sua}.

We will not attempt to directly map the matter power spectrum at the scales observable by 21-cm data ($k\sim0.1-1\,\Mpcinv$).
Instead, we will use the large-scale 21-cm fluctuations as a tracer of small-scale structure formation, mirroring our analysis for the global signal.
Throughout this section we will obtain theoretical results with the same {\tt 21cmvFAST} simulations as in the previous section.
For convenience we define the amplitude of 21-cm fluctuations as
\be
\Delta^2_{21}(k_{21}) = \dfrac{k_{21}^3}{2\pi^2}P_{21}(k_{21}),
\ee
where $P_{21}$ is the 21-cm power spectrum, although we will refer to $\Delta^2_{21}(k_{21})$ as the 21-cm power spectrum unless confusion can arise.
Here, and throughout, we will denote the wavenumber of the 21-cm fluctuations by $k_{21}$, in order to avoid confusion with the matter power spectrum modes $k$, and we will often refer to measurements of our observable---the 21-cm power spectrum---as 21-cm fluctuations, for the same reason.

\begin{figure}[t!]
	\centering
	\includegraphics[width=0.46\textwidth]{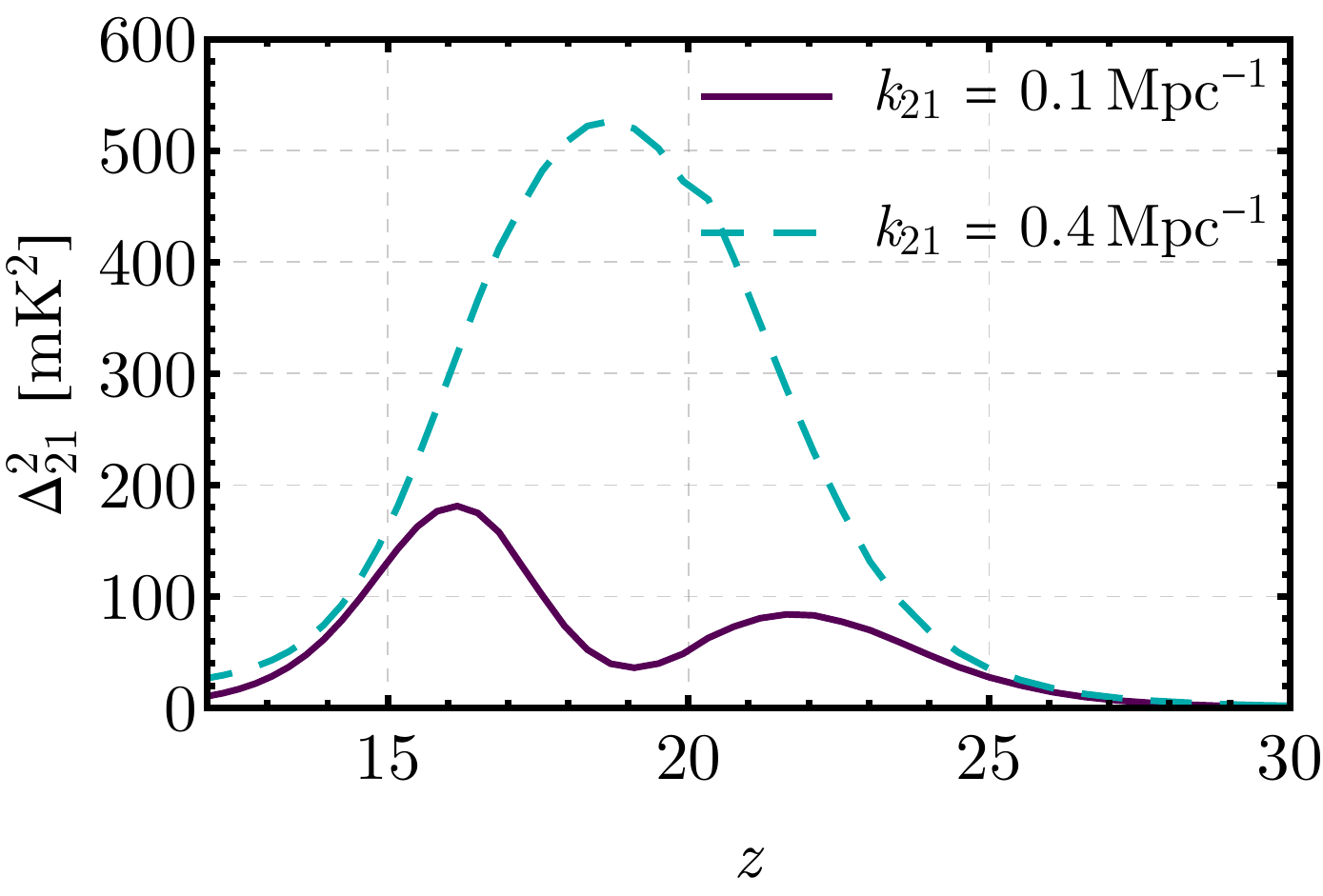}
	\caption{Amplitude of 21-cm fluctuations as a function of redshift, for two characteristic wavenumbers, $k_{21}=0.1\Mpcinv$ and $k_{21}=0.4\Mpcinv$.
	The 21-cm power at both wavenumbers traces the overall formation of the first stars, starting at $z\sim 25$, transitioning from the LCE to the EoH at $z\sim 19$, and nearly vanishing after the universe is fully heated.
	}	
	\label{fig:P21fid}
\end{figure}

We show our fiducial 21-cm power spectrum at two different wavenumbers in Fig.~\ref{fig:P21fid}.
Both the large-scale ($k_{21}=0.1$ $\rm Mpc^{-1}$) and small-scale ($k_{21}=0.4$   $\rm Mpc^{-1}$) 21-cm fluctuations follow the same overall trend as the global signal, with small changes.
First, at $z\sim 25$, the LCE begins and, due to the anisotropic nature of the first stellar formation, large 21-cm anisotropies are generated and the power spectrum grows at all scales.
The small-scale ($k_{21}=0.4$  $\rm Mpc^{-1}$) power keeps growing until the EoH, after which it slowly decreases until at the end of our simulations, where the universe is fully heated.
The large-scale ($k_{21}=0.1\,\rm Mpc^{-1}$) power, however, starts decreasing during the LCE, nearly vanishing at the transition between the LCE and the EoH (at $z\sim 19$).
This is because the power at these scales is dominated by the VAOs, which have opposite effects during the LCE and EoH, as large velocities produce fewer stars, reducing the 21-cm absorption during the LCE but increasing it during the EoH (for more details see Ref.~\cite{Munoz:2019rhi}).

\begin{figure}[hbtp!]
	\centering
	\includegraphics[width=0.46\textwidth]{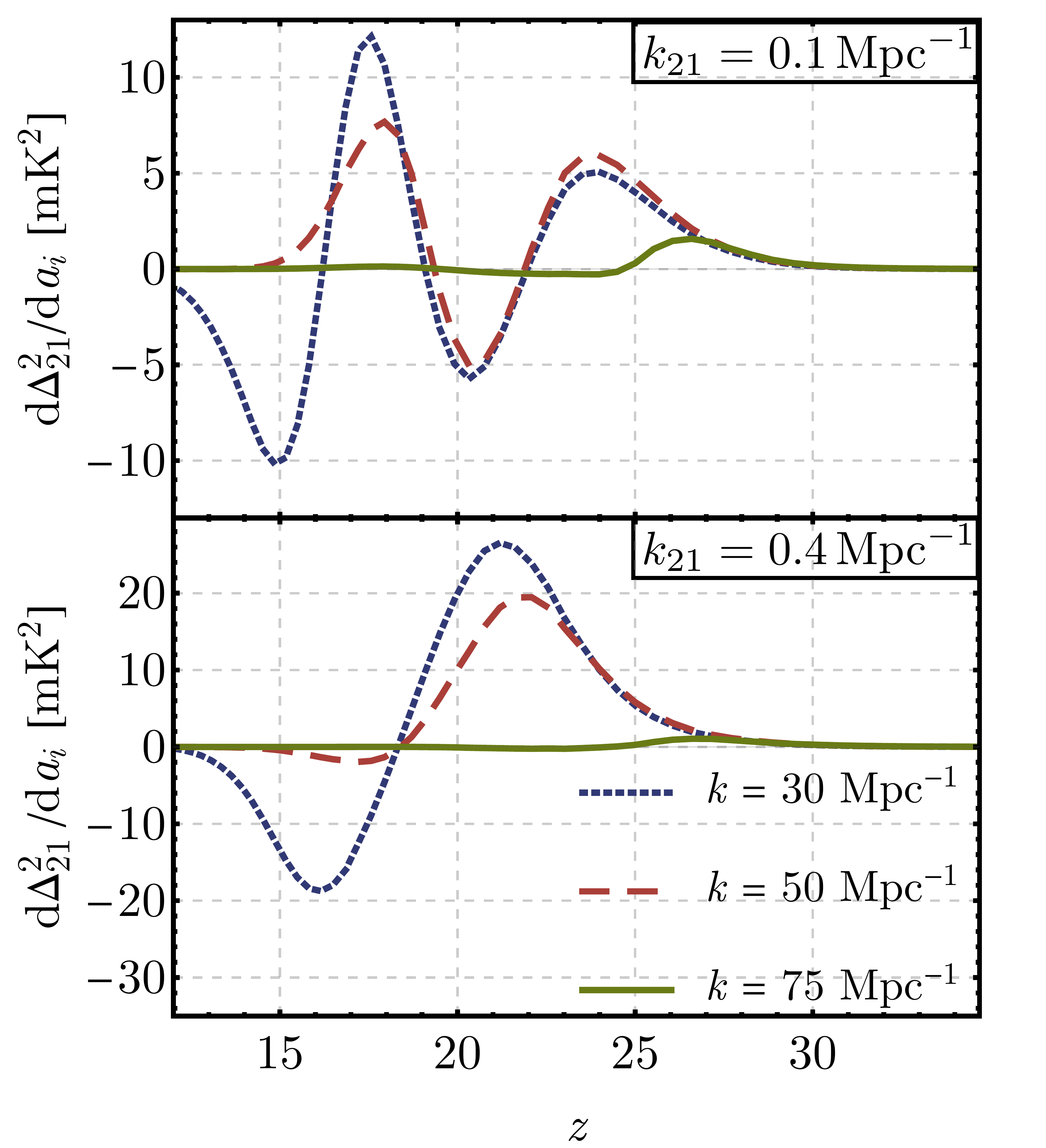}
	\caption{Derivative of the amplitude of 21-cm fluctuations with respect to the bin amplitudes for our three chosen modes, as a function of redshift $z$.
	The top panel shows the result for a large-scale 21-cm wavenumber ($k_{21}$), whereas the bottom panel shows a smaller-scale one.
	As before, large-$k$ matter fluctuations only affect the signal at high redshifts, and these derivatives change sign whenever the fiducial $\Delta^2_{21}(k_{21})$ has a local extremum.
	}	
	\label{fig:dP21bins}
\end{figure}

 \subsection{Small-Scale Matter Power Spectrum}
 
We begin by studying how well the matter power spectrum can be constrained, by varying its amplitude on different $k$-bins, as we did in Sec.~\ref{sec:global}.
 
We show in Fig. \ref{fig:dP21bins} the derivatives of two 21-cm fluctuation modes, with $k_{21}=0.1\Mpcinv$ and $0.4\Mpcinv$, with respect to the amplitude of the matter power spectrum at the three illustrative modes as before.
As in the global-signal case, we see that the lowest $k$-bin changes the signal across all redshifts, whereas smaller scales (larger $k$) preferentially affect larger redshifts, where smaller haloes were more commonly forming stars.

The governing principle that sets the shape of the derivatives in Fig.~\ref{fig:dP21bins} is that more matter power translates into more stars, and thus a faster evolution of the signal.
This is why  the derivative at $k_{21}=0.4\,\rm Mpc^{-1}$ grows during the LCE and decreases during the EoH, as the overall signal at this wavenumber peaks at $z=20$ (c.f.~Fig.~\ref{fig:P21fid}).
The evolution is slightly more complicated for the larger-scale mode with $k_{21}=0.1\,\rm Mpc^{-1}$, as the amplitude of the 21-cm fluctuations has two peaks, one during the LCE at $z=22$, and another one during the EoH at $z=16$.
Thus, the derivative at this wavenumber changes sign twice.
As in the global-signal case, the differences between different matter power-spectrum modes will allow us to distinguish them, 
although now we will have spatial information in the form of different $k_{21}$ modes.

 \subsection{Astrophysical Parameters}
 
As was the case for the global signal, changing the amplitude of the matter power spectrum (especially at low $k$) will be very degenerate with changing the astrophysical parameters.
We now quantify this.

\begin{figure}[hbtp!]
	\centering
	\includegraphics[width=0.46\textwidth]{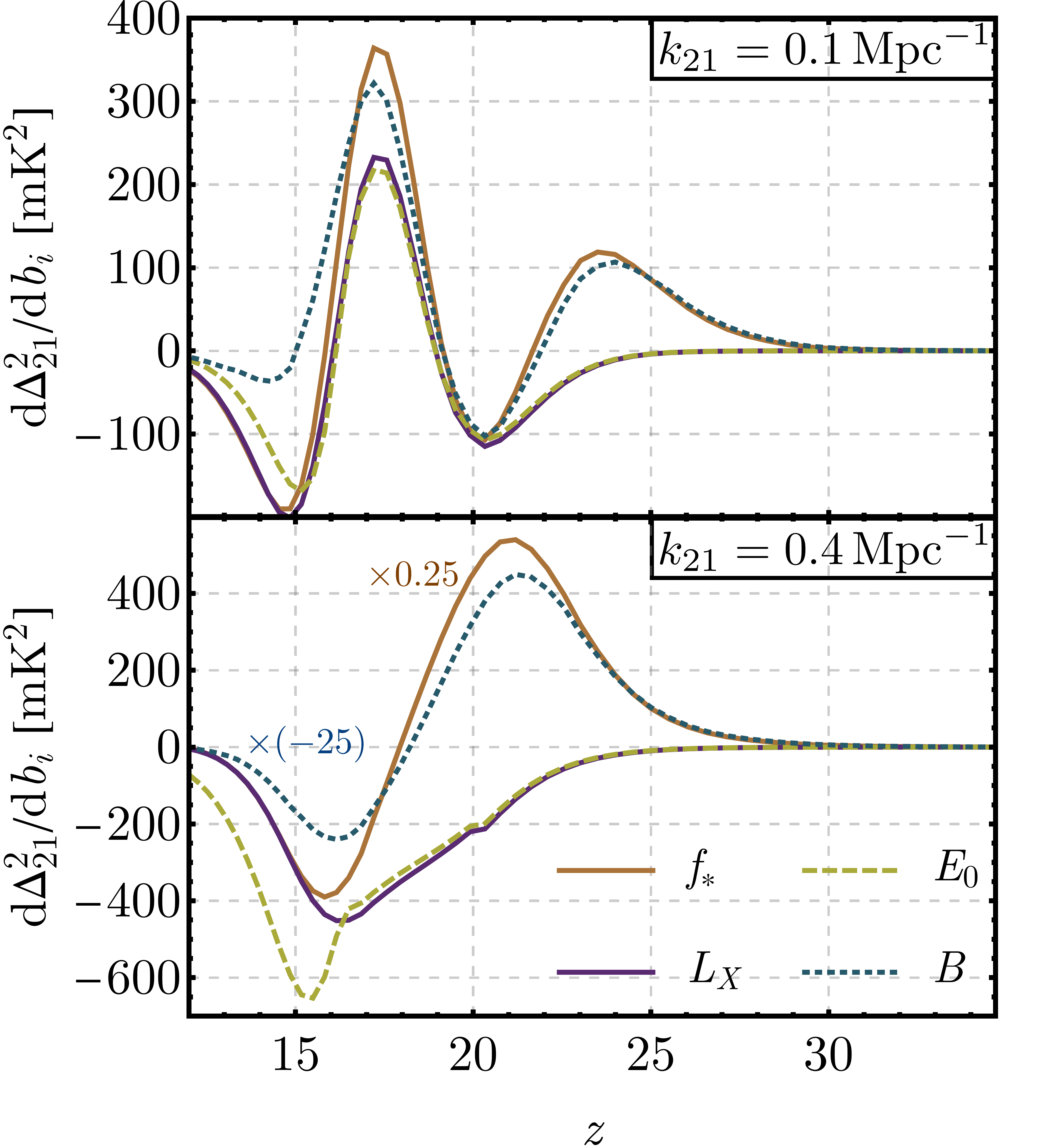}
	\caption{Derivative of the 21-cm power spectrum with respect to the astrophysical parameters, at two representative wavenumbers $k_{21}$, as a function of redshift $z$.
	The addition of spatial information (in the form of different wavenumbers $k_{21}$) allow us to break degeneracies between parameters.
	}	
	\label{fig:dP21astro}
\end{figure}
 
We consider the same astrophysical parameters as in Sec.~\ref{sec:global},
Notice, however, that we will now have more information on the effect of each parameter, as we can probe not only the redshift evolution of the 21-cm signal, but also its scale-dependence.
This is clear in Fig.~\ref{fig:dP21astro}, where we show the derivatives with respect to each astrophysical parameter for the same two 21-cm wavenumbers as before.
Parameters that had almost identical derivatives in global-signal studies, such as $f_*$ and $B$, can be better distinguished using 21-cm fluctuations.

 \subsection{Foregrounds}

In the global-signal analysis carried out in Sec.~\ref{sec:global}, foregrounds cannot be separated from the cosmological signal, and thus have to be fitted simultaneously.
In contrast, the angular information in power-spectrum data allows for a separation of foreground- and cosmology-dominated regions, as foregrounds are restricted to the wedge.
In terms of line-of-sight (LoS) wavenumbers $k_{||}$, and perpendicular ones $k_\perp$, the wedge is defined to contain modes with
\be
k_{||} \leq a(z) + b(z) k_\perp,
\label{eq:wedge}
\ee
where the values of $a$ and $b$ define the extent of the wedge~\cite{Pober2013,Pober:2013jna}.

Within the wedge, foregrounds are much larger in size than the expected 21-cm signal, so this region is discarded for cosmology analyses.
Given our ignorance about the wedge, we will show results under three foreground assumptions, mirroring those of Ref.~\cite{Munoz:2019rhi} (based on Refs.~\cite{Pober2013,Pober:2013jna}), with values of $a=\{0,0.05,0.1\}\, h\Mpcinv$ for optimistic, moderate, and pessimistic foregrounds, respectively; and $b$ given by the horizon limit in all cases except optimistic, where it is smaller by a factor of $\sin(\theta_b/2)$, where $\theta_b$ is the beam full-width half maximum.
We encourage the reader to visit Refs.~\cite{Pober2013,Pober:2013jna} for more details.

\begin{figure}[hbtp!]
	\centering
	\includegraphics[width=0.46\textwidth]{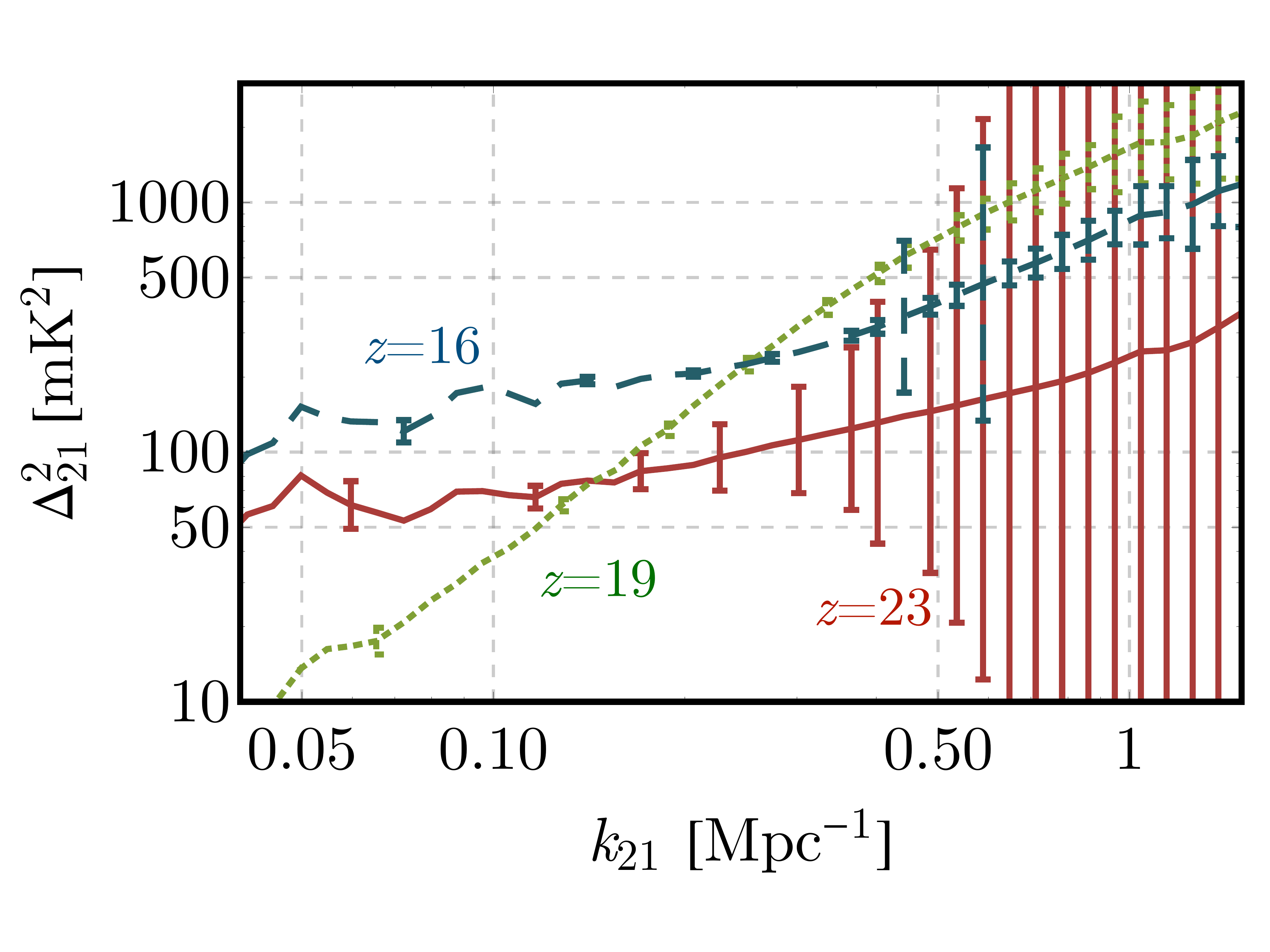}
	\caption{Fiducial 21-cm power spectrum as a function of wavenumber $k_{21}$, along with our forecasted HERA error bars, at three redshifts.
	These are chosen to roughly correspond to the half points of the LCE ($z=23$, in red), the EoH ($z=16$, in blue dashed), and the transition between these two eras ($z=19$, in green dotted).
	The curves are obtained with a modified version of {\tt 21cmvFAST}, and the error bars correspond to the moderate-foreground case explained in the text, using {\tt 21cmSense}.
	We assume three years of HERA data, and the wavenumbers that do not have error bars cannot be measured at any precision.
	}	
	\label{fig:P21errors}
\end{figure}

 \subsection{Results}

We now use the derivatives defined above to determine how well future 21-cm experiments can measure the matter power spectrum and astrophysical parameters.

We calculate the expected noise for HERA with the publicly available code {\tt 21cmSense}~\cite{Pober2013,Pober:2013jna}\footnote{\url{https://github.com/jpober/21cmSense}}.
We consider the redshift range $z=14-27$ ($\nu=51-95$ MHz), which we divide in 14 bands, each 4 MHz wide.
We take three years of observations (540 days, with 8 hours per day), and throughout this section we assume a sky temperature given by $T_{\rm sky} (\nu) = 100 + 120 (\nu/150\,\rm MHz)^{-2.55}$ K, as in Ref.~\cite{DeBoer:2016tnn}, when computing power-spectrum noise.

We show in Fig.~\ref{fig:P21errors} the resulting error bars on top of our fiducial CDM power spectrum for three redshifts, chosen to illustrate some of the landmarks of the cosmic-dawn era.
We show the power spectrum at $z=19$, during the transition between the LCE and the EoH, where the global 21-cm signal reached its minimum.
In this case there is only power at small scales (large $k_{21}$).
We also plot two additional redshifts, halfway through the EoH and the LCE, corresponding to $z=16$ and $z=23$, respectively, where the global 21-cm signal has half of its deepest absorption value.
In both of these cases VAOs can be discerned in the simulation data~\cite{Dalal:2010yt,Visbal:2012aw,Munoz:2019rhi}, more distinctly perhaps at $z=16$.
We also show in Fig.~\ref{fig:P21errors} the forecasted HERA errors for each case.
We note, in passing, that higher-$z$ bins in Fig.~\ref{fig:P21errors} can reach lower-$k_{21}$ modes, as our fixed bandwidth of 4 MHz corresponds to a wider redshift range at higher redshifts (lower frequencies), and thus the minimum $k_{||}$ observable is larger.

\begin{figure}[hbtp!]
	\centering
	\includegraphics[width=0.46\textwidth]{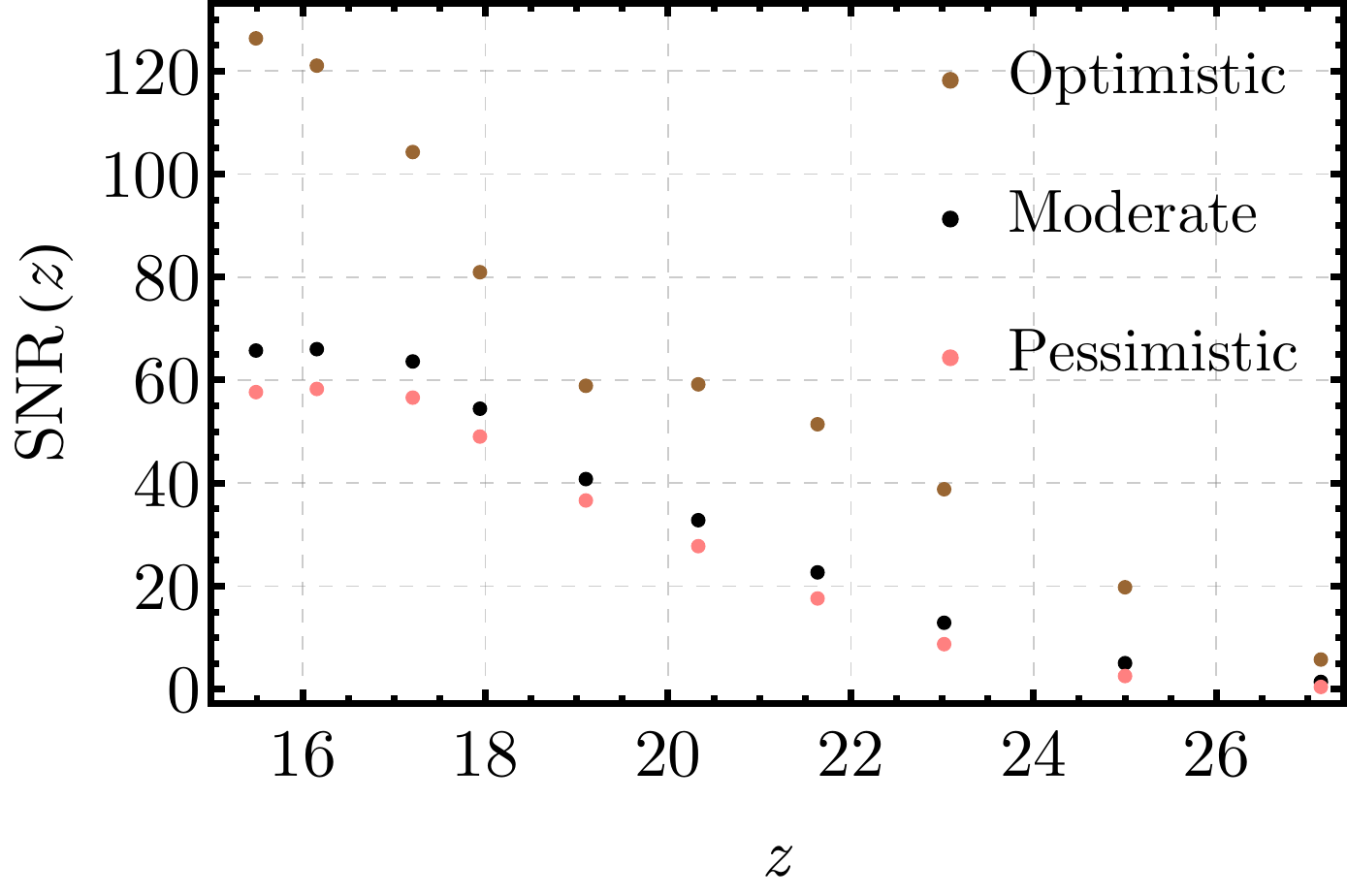}
	\caption{Signal-to-noise ratio (SNR) of our fiducial 21-cm power spectrum as a function of redshift, for each of the three foreground scenarios considered, in all cases with three years of HERA data.
	}
	\label{fig:SNRP21}
\end{figure}

Let us start by computing the signal-to-noise ratio SNR at a single band centered around frequency $\nu$, defined to be
\be
{\rm SNR} (\nu) = \left( \sum_{k_{21}-{\rm bins}} \left[ \dfrac{\Delta^2_{21}(\nu,k_{21})}{\sigma(\Delta^2_{21})} \right]^2 \right)^{1/2},
\ee
which is only a function of redshift, as we integrate over all $k_{21}$ bins but not over frequencies.
Here $\sigma(\Delta^2_{21})$ are the errors obtained from {\tt 21cmSense}.
As mentioned in the previous subsection, we will consider three different foreground scenarios, for which we show the SNR as a function of redshift in Fig.~\ref{fig:SNRP21}.
Note that the foregrounds strength not only determines the size of the error bars at each wavenumber $k_{21}$, but also which values of $k_{21}$ can be observed to begin with.
This is why the difference between the optimistic and moderate cases is larger than between moderate and pessimistic, as the lower $b(z)$ in Eq.~\eqref{eq:wedge} allows for many more modes to be observed.
The total SNR of our fiducial case, obtained by summing in quadrature over redshifts, is SNR$=\{120, 140, 240\}$ for pessimistic, moderate, and optimistic foregrounds.

As in the global-signal case, the SNR is a good indication of how well the signal can be measured.
Nonetheless, we need to know how correlated different parameters are.
Here we construct our Fisher matrix as~\cite{Liu:2015gaa, Mao:2008ug}
\be
F_{\alpha \beta} = \sum_{k_{21}-{\rm bins}}  \sum_i \dfrac{\partial \Delta^2_{21}(\nu_i,k_{21})}{\partial \theta_\alpha} \dfrac{\Delta^2_{21}(\nu_i, k_{21})}{\partial \theta_\beta} \sigma^{-2}(\Delta^2_{21}),
\ee
where we again assume that the covariance matrix is diagonal in redshift and wavenumbers (ignoring trispectrum contributions~\cite{Shaw:2019qin}), so $\sigma(\Delta^2_{21})$ is the error obtained with {\tt 21cmSense} at each frequency $\nu_i$ and  $k_{21}$ bin.

We begin by finding how well HERA could measure the astrophysical parameters within CDM.
Fig.~\ref{fig:Astro_ellipses} shows the confidence ellipses for our four astrophysical parameters under moderate foregrounds.
We see that, while the SNR is significantly lower than our global-signal forecast (SNR$^{\rm Fluct.}=140$ versus SNR$^{\rm GS}=560$), the absence of foreground parameters to marginalize, added to the additional angular information, allows 21-cm fluctuations to more-precisely measure the astrophysical parameters than the global signal.
The only parameter with similar constraints for both observables is $E_0$, which was not very degenerate in global-signal studies.
Moreover, our results align with those of Ref.~\cite{Greig:2017jdj}, although we consider a different fiducial model and three years of HERA data, versus only one, resulting in better constraints.
We forecast uncertainties of $\sigma(f_*)=0.01$, $\sigma(\log_{10}(L_X)) =0.03$, $\sigma (E_0)=0.02$ keV, and $\sigma(B) = 1.4$.

In order to measure the matter power spectrum we follow an analysis similar to the global-signal case, packing wavenumbers into wide bins, with amplitudes $\mathbf {\tilde a}$.
In this section, however, we will deviate in two regards.
First, we will divide the $k$-range observable in four bins, as power-spectrum measurements can more easily distinguish between adjacent---and thus correlated---bins.
We keep the same lowest- and largest-$k$ bins, but split the middle bin into two, covering the range $k=(38-60)\Mpcinv$ and $k=(60-80)\Mpcinv$, respectively.
Second, we will not impose a prior on the astrophysical parameters ($f_*,B,$ and $E_0$), as the additional information in the fluctuations (in the form of long- and short-wavelength modes) allows us to disentangle them better from each other, and from the bin amplitudes.

In our baseline case of moderate foregrounds, we forecast uncertainties of $\sigma(a_1) = 0.05$, $\sigma(a_2)=0.2$, and  $\sigma(a_3)=0.3$ for the amplitude of the first three bins, and again the highest-$k$ bin cannot be measured at any precision, with $\sigma(a_4)=9 \gg 1$.
All of these uncertainties are significantly better than with global-signal measurements, especially for the first bin, due to the extra information contained in large and small wavenumbers.
This results into stronger constraints of the matter power spectrum, as shown in Fig.~\ref{fig:Pmz0_bins}, where a HERA-like experiment can outperform an EDGES-like global signal detection by a factor of $\sim 5$, given the same bin selection.

\begin{table}[hbtp!]
	\noindent\begin{tabular}{ l | c  c  c }	
		\multicolumn{1}{c}{} 	&\multicolumn{3}{c}{Foregrounds}\\
		\cline{2-4}
		Wavenumber range  &  Optimistic  & Moderate &  Pessimistic \\
		\hline
		$k=(3-38)\Mpcinv$ & 0.53\% & 1.2\%& 1.7\%  \\
		
		$k=(38-60)\Mpcinv$  & 5.5\%& 16\%& 23\% \\
		
		$k=(60-80)\Mpcinv$  & 7.8\%& 29\%& 83\% \\
\hline
\hline		
	\end{tabular}
	\caption{Projected 1-$\sigma$ uncertainties on the amplitude of the matter power spectrum integrated over different wavenumber ranges, under the three foregrounds assumptions described in the text.
	These results have been obtained by marginalizing over astrophysical parameters.}
	\label{tab:errorbins}
\end{table}

We show the results for the other foreground assumptions in Table~\ref{tab:errorbins}, where in all cases it should be possible to measure the matter power spectrum with precision ranging from percent-level for the first bin ($k=(3-38)\Mpcinv$) to tens of percent for higher $k$.
Further sub-dividing the $k$-range into finer bins would result in significantly worse constraints, due to the degeneracies between adjacent modes.
The number of bins chosen (three for global signal and four for power spectrum) allows for $\sim 10\%$-level precision measurements of the matter fluctuations.

\section{Principal Components}
\label{sec:PCs}

Our analysis thus far has consisted of simply varying the amplitude of the matter power spectrum in bins centered around different wavenumbers, and estimating how well the amplitude of each bin can be measured.
These bins are, however, highly correlated with each other and with the astrophysical parameters, resulting in non-optimal constraints.
We will now improve the analysis above by finding the principal components (PCs) of the matter power spectrum for our two observables.

\subsection{Method}

We aim to decompose the matter power spectrum as
\be
\Delta^2_m = \Delta^2_{m,\rm fid} \left[1 +\sum_a m_a {\rm PC}_a(k) \right],
\label{eq:PmatterPCs}
\ee
where the functions PC$_a(k)$ form an orthonormal set.
These functions can be built from linear combinations of the $f_i$ introduced in Eq.~\eqref{eq:Pmatter}.
Here we will use the finer binning of the matter power spectrum described in Sec.~\ref{sec:model}, dividing the range of interest ($k=(20-108)\,\rm Mpc^{-1}$) in 40 bins, logarithmically spaced.
We begin by computing the Fisher matrix for our two observables (the 21-cm global signal and fluctuations), including all astrophysical and foreground parameters (when relevant), as well as these 40 bin amplitudes ($\mathbf a$).

We can write our Fisher matrix as
\be 
F_{\alpha \beta} =
\begin{pmatrix}
	F_{ij} & F_\times\\
	F_\times& F_{A B}\\
\end{pmatrix},
\ee
where $F_{ij}$ and $F_{A B}$ are the matrix blocks corresponding to the bin amplitudes and nuisance parameters, respectively, and $F_\times$ contains the cross terms.
We build the principal components by diagonalizing the $(i,j)$ block of the covariance (inverse-Fisher) matrix, which corresponds to the (marginalized) bin amplitudes.
For computational convenience we compute the ``degraded Fisher matrix" for our bin amplitudes~\cite{Leach:2005av},
\be
F_{ij}^{\rm deg} = F_{ij} - F_\times^T F_{AB} F_\times,
\ee
which accounts for marginalization over the rest of parameters, as $C_{ij} = [(F^{\rm deg})^{-1}]_{ij}$, and diagonalize this matrix instead.

\begin{figure}[b!]
	\centering
	\includegraphics[width=0.46\textwidth]{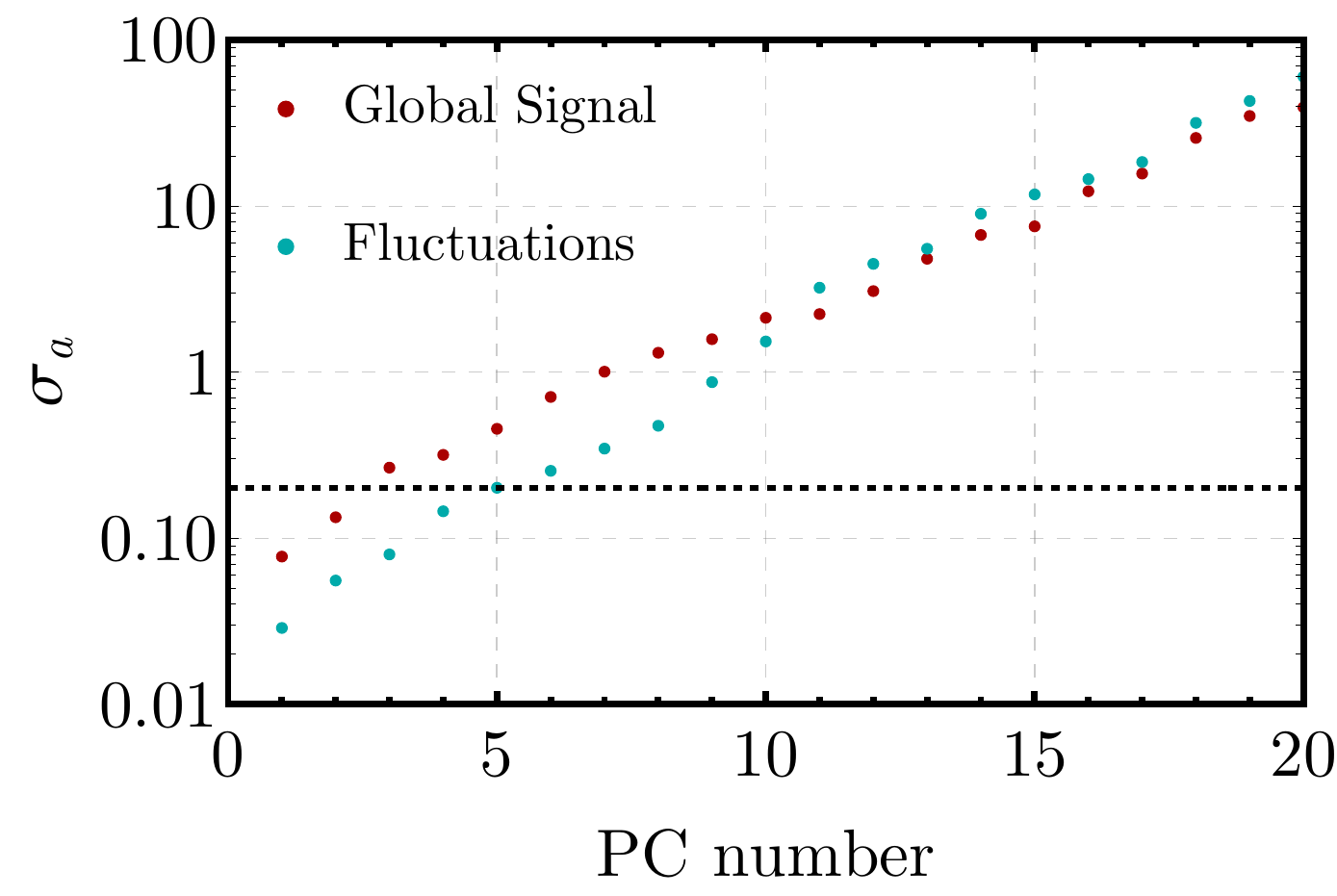}
	\caption{Forecasted uncertainties in each PC amplitude for the 21-cm global-signal (red) and fluctuations (teal) cases outlined in the text, where in the latter we have assumed moderate foregrounds.
	Setting a threshold at a SNR$=5$ ($\sigma_a<0.2$), the global signal can detect 2 PCs, whereas for the fluctuations that number is 4 (as the fifth is slightly beyond the threshold).
	}	
	\label{fig:PCerrors}
\end{figure}

We define each PC by constructing linear combinations of our bins:
\be
{\rm PC}_a(k) = \sqrt{N_{\rm bins}} \sum_i S_a^i f_i(k),
\label{eq:PCs}
\ee
where the $S_a$ are the eigenvectors of $F_{ij}^{\rm deg}$, and the prefactor of the number $N_{\rm bins}$ of bins in this equation normalizes the PCs, so that
\be
\int d\log k \, {\rm PC}_a(k)  {\rm PC}_b(k) = \delta_{a b}
\ee
over our chosen $k$-range.
The PCs are thus built to be orthonormal, and their expected uncertainties are given by~\cite{Hu:2003vp,Paykari:2009ac,Dvorkin:2010dn,Dvorkin:2011ui}
\be
\sigma_a = ( N_{\rm bins} \lambda_a)^{-1/2},
\ee
where  $\lambda_a$ are the eigenvalues of $F_{ij}^{\rm deg}$ ordered upwards.
We show these errorbars in Fig.~\ref{fig:PCerrors}, both for our global signal and power-spectrum analyses (assuming moderate foregrounds), where we can measure (at SNR $>5$, i.e., $\sigma_a<0.2$) 2 and 4 PCs respectively, as the uncertainty of lower-SNR PCs rapidly grows.
As a note, we include (broad) priors for our astrophysical parameters for the global-signal, but not for the fluctuations, as in Sec.~\ref{sec:fluctuations}.
  
   \begin{figure}[t!]
	\centering
	\includegraphics[width=0.46\textwidth]{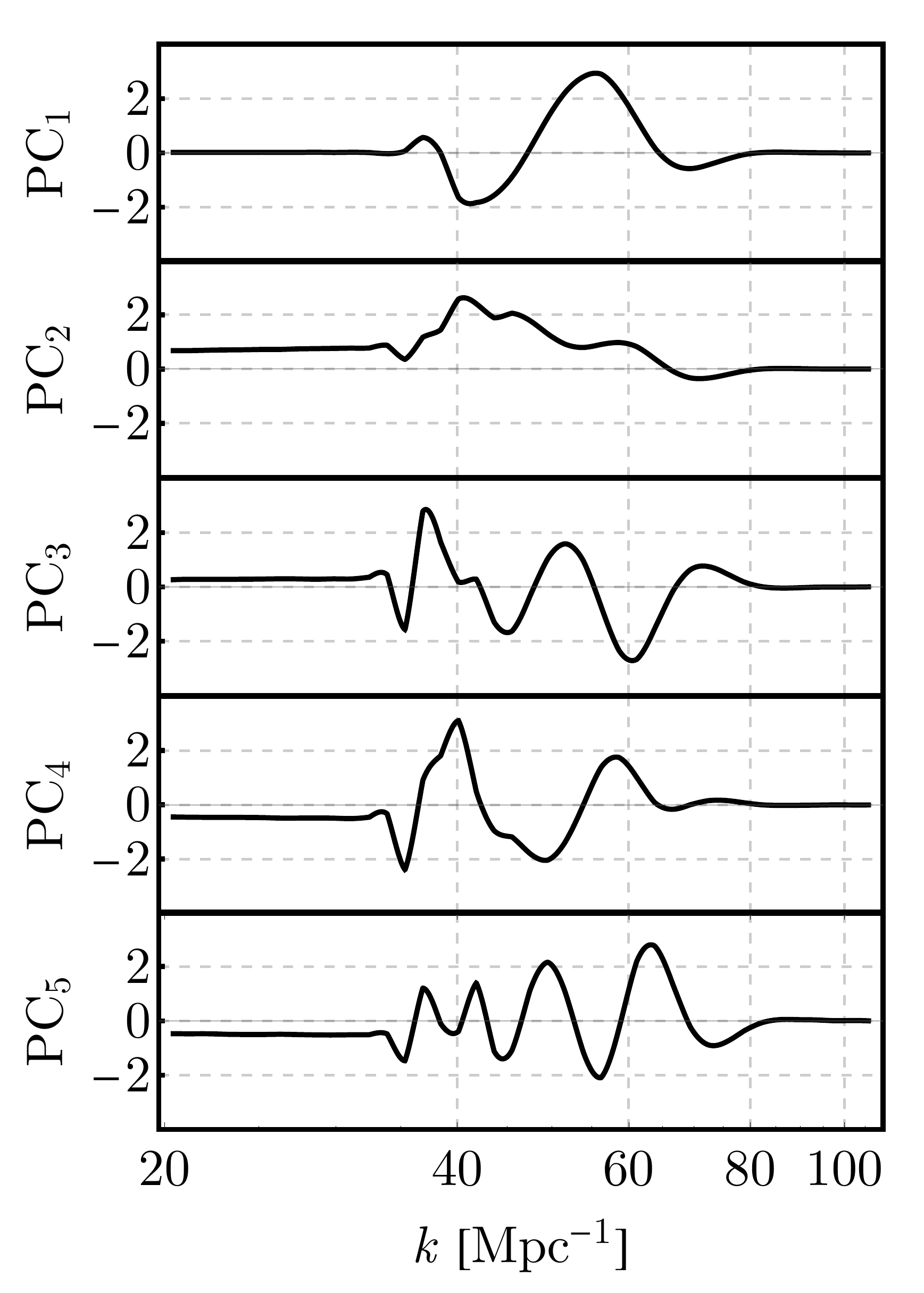}
	\caption{The first five PCs of the matter power spectrum for our 21-cm global-signal analysis, which can all be measured with SNR $>1$.
	The majority of the support of the PCs is for $k\sim40-80\Mpcinv$, as the effect of those modes is sizable and can be distinguished from nuisance parameters.
	}	
	\label{fig:PCshapes_GS}
\end{figure}
    
   \begin{figure}[t!]
   	\centering
   	\includegraphics[width=0.46\textwidth]{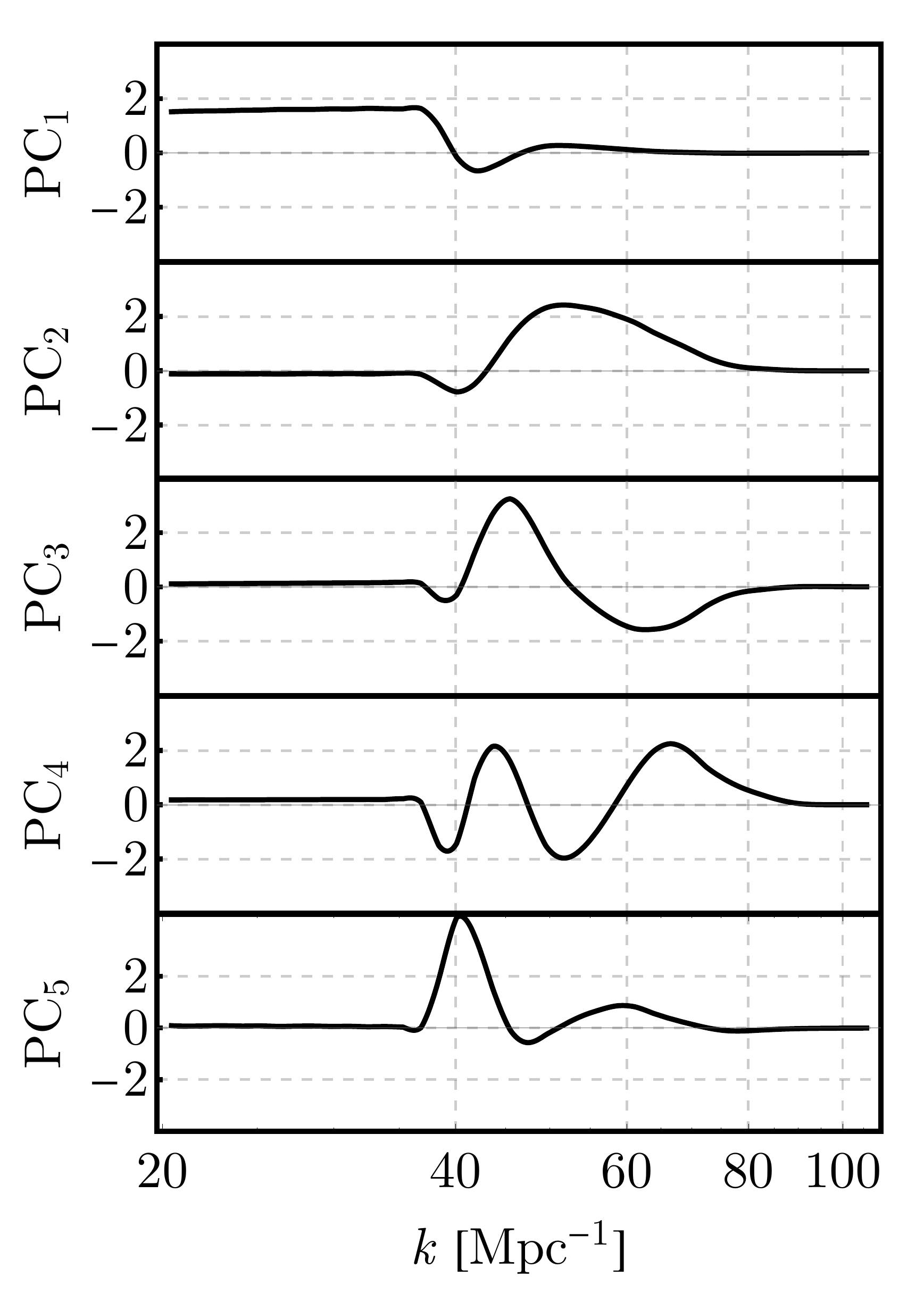}
   	\caption{Same as Fig.~\ref{fig:PCshapes_GS} but for the 21-cm fluctuations, under moderate foregrounds.
   	In addition to peaking at $k\sim40-80\Mpcinv$, some of these PCs show support for $k<40\Mpcinv$, as the 21-cm fluctuations can distinguish those wavenumbers from astrophysical parameters to some degree.
   	}	
   	\label{fig:PCshapes_PS}
   \end{figure}
 
\subsection{Results}
   
The shape of the first five PCs, determined with Eq.~\eqref{eq:PCs} by splining through the basis functions $f_i(k)$, is shown in Figs.~\ref{fig:PCshapes_GS} and~\ref{fig:PCshapes_PS} for our two observables.
There are two important features of these PCs worth mentioning.

First, both for the 21-cm global signal and the fluctuations, the first PCs are peaked around $k\sim 50\Mpcinv$, indicating that this is the $k$-range where the power spectrum can be best probed.
That is because for lower wavenumbers the effects are hard to distinguish from an overall normalization (and are thus degenerate with the nuisance parameters and with each other), whereas higher wavenumbers only affect high-redshift 21-cm signals, where foregounds are stronger and the SNR is lower.
As expected, higher-order PCs typically show more oscillations around the best-measured wavenumbers, capturing additional features beyond the amplitude of the power spectrum at a particular $k$.

Second, as discussed in Sec.~\ref{sec:global}, and shown in Fig.~\ref{fig:Corrfirstbin}, all wavenumbers with $k<k_{\rm atom}\sim 38\,\Mpcinv$ have nearly unity correlation.
This is clear in Fig.~\ref{fig:PCshapes_GS}, where those modes have roughly identical weights in our PCs, as they have a very similar effect in both 21-cm observables.
While the global-signal PCs show little weight at lower $k$, due to the degeneracies with astrophysical parameters, the situation is slightly more optimistic for the 21-cm fluctuations, where some of the PCs show tails extending to low $k$, as the bin amplitudes can be separated from $f_*$ and $B$.

These two features inspired our choice of $k$ ranges in the previous sections, as the bins shown in Fig.~\ref{fig:Pmz0_bins} roughly correspond to the $k<k_{\rm atom}$ regime, and the bump of the first few PCs at $k\sim 50\Mpcinv$.
The 21-cm fluctuations (with 4 PCs measurable at SNR $>5$) can reach higher wavenumbers than the global signal (with only 2 PCs),
but both are sensitive to deviations from the standard CDM model at the 10\% level through their projection on the PCs, showcasing the power of the 21-cm line to measure the small-scale distribution of matter.

\subsection{Constraints on Dark-Matter Models}

One of the major benefits of developing the  model-agnostic PCs shown above is that they can be used for constraining any model with a small-scale power spectrum that deviates from our fiducial CDM.
We will now use the PCs to forecast constraints for three non-CDM (nCDM) models.

Given a nCDM model, we can write its power spectrum as
\be
\Delta^2_m(k) = T_{\rm nCDM}^2(k) \Delta^2_{m,\rm fid} (k),
\ee
where the transfer function $T_{\rm nCDM}$ encodes the deviation from our fiducial CDM model.
We will obtain constraints by projecting the deviation from CDM onto the PCs for our two observables.
We calculate the SNR of the $a$-th PC to this deviation as
\be
{\rm SNR}_a = \int \dfrac{d \log k }{\Delta \log k}\, \dfrac{{\rm PC}_a(k)}{\sigma_a} \left[1-T_{\rm nCDM}^2(k)\right],
\ee
and when adding a number $N_{\rm PC}$ of PCs we can simply add their SNRs in quadrature, as they are orthonormal.
Then, the SNR is zero for CDM, and larger values imply a preference for the nCDM model over CDM.

We will test three popular nCDM models, posited to resolve different small-scale puzzles in CDM.
For concreteness we will find the point at which each model crosses SNR $=2$, chosen to roughly correspond to a 95\% C.L.~rejection, as a metric of how small of a departure from CDM 
we can detect.
Our results are summarized in Table~\ref{tab:limits}.

The first model we consider is warm dark matter (WDM), where we can use the simple analytic approximation for the suppression from~\cite{Bode:2000gq}, introduced in Sec.~\ref{sec:global}.
As an example, we note that for the current 95\% C.L.~limit of $m_{\rm WDM}=5$ keV~\cite{Irsic:2017ixq}, we forecast that the first two global signal PCs have SNR$_a^{\rm GS}=\{2.1,3.6\}$, and the rest of PCs have nearly zero SNR, which yields a total SNR$^{\rm GS}=4.1$.
For the fluctuations those numbers are SNR$_a^{\rm Fluct.}=\{5.6,4.7\}$, and thus the total is SNR$^{\rm Fluct.}=7.1$.
In this simple case, we have tested our PC method against a direct Fisher-matrix forecast of the amplitude of the WDM suppression, marginalizing over all astrophysical parameters, and found the same values of SNR within 2\%.
This validates our PC approach.
Using our SNR $=2$ criterion from above, we find that we can probe up to
$m_{\rm WDM}=8.3$ keV with the 21-cm global signal, and 14 keV with the fluctuations.
This is significantly more powerful than current limits from the Lyman-$\alpha$ forest~\cite{Irsic:2017ixq,Viel:2005qj,Baur:2015jsy,Seljak:2006qw}.

\begin{table}[]
	\begin{tabular}{c | c | c | c | c}
		Model & Parameter      & Current & Global Signal & Fluctuations \\
		\hline 
		WDM   & $m_{\rm WDM}$ [keV] & $> 5.3$     & 8.3           &      14        \\
		FDM   & $m_a$ [$10^{-21}$eV]    & $>2.0 $   & $24$       &    $48$          \\
		ETHOS  & $\mathfrak{a}_4$ [Mpc$^{-1}$]          & $< 480$      & 4          &      1        \\       
	\end{tabular}

\caption{Limits on the parameters of different non-CDM models.
For warm DM (WDM) the relevant parameter is its mass, where the current lower limit is from Ref.~\cite{Irsic:2017ixq}.
For fuzzy DM (FDM) the parameter is the axion mass, with the current lower limit from Ref.~\cite{Irsic:2017yje}.
For ETHOS the parameter is the amplitude $\mathfrak{a}_4$ of the Yukawa coupling between DM and DR (normalized at $z=10^7$), where the current best upper limit is obtained from Ref.~\cite{Archidiacono:2017slj} (see text for details and note that, as opposed to the cases of WDM and FDM, larger values of the parameter $\mathfrak {a}_4$ deviate further from CDM).
In all cases we assumed that each candidate composes all of the DM.
\label{tab:limits}}
\end{table}

The second DM model we consider is that of axion-like particles, also known as fuzzy DM (FDM)~\cite{Hu:2000ke,Hui:2016ltb,Marsh:2015xka}.
In that case the transfer function can be well approximated as~\cite{Hu:2000ke}
\be
T_{\rm FDM} (k) = \dfrac{\cos (x^3)}{1+x^8},
\ee
with $x=1.61 m_{a,22}^{1/18} \, k/k_{J,\rm eq}$ and $k_{J,\rm eq}=9 m_{a,22}^{1/2}$ Mpc$^{-1}$,
where $m_{a,22}$ is the mass $m_a$ of the axion-like particle in units of $10^{-22}$ eV.
The suppression in this model is more sudden than for WDM, so 21-cm can significantly improve upon Lyman-$\alpha$ limits, as they reach smaller scales albeit at worse precision.
Indeed, as we show in Table~\ref{tab:limits}, the 21-cm global signal and fluctuations can probe FDM masses as large as $m_{\rm FDM}=2.4$ and $4.8\times 10^{-20}$ eV, respectively, whereas the best current constraints from the Lyman-$\alpha$ forest are about an order of magnitude worse: $m_{\rm FDM}=2\times 10^{-21}$ eV~\cite{Irsic:2017yje}.
This is not the only way to search for FDM, however, as its quantum nature allows for small fluctuations to gravitationally heat (and thus disrupt) star clusters~\cite{Marsh:2018zyw}, which rules out FDM with $m_{\rm FDM} = 10^{-20} - 10^{-19}$ eV.
This leaves a window around $m_{\rm FDM} \sim 10^{-20}$ eV that the 21-cm line can probe (see also~\cite{Lidz:2018fqo,Schneider:2018xba}).

The last case we study is that of DM interacting with a dark-radiation (DR) bath, also introduced in Sec.~\ref{sec:global}.
As in that section, we focus on Yukawa-like interactions with a massive mediator (corresponding to $n=4$ in ETHOS~\cite{Cyr-Racine:2015ihg,Vogelsberger:2015gpr}).
The relevant parameter is, then, the constant $\mathfrak{a}_4$, with units of inverse length, which determines the size of the DM-DR interactions (see Eq.~\eqref{eq:ETHOS_4} above).
We find that the 21-cm global signal can probe opacities down to $\mathfrak{a}_4=4 \Mpcinv$, which is improved to $\mathfrak{a}_4=1 \Mpcinv$ by the 21-cm fluctuations.
In order to put these numbers in context, we will compare them with current limits coming from CMB and Lyman-$\alpha$-forest data~\cite{Archidiacono:2019wdp,Bose:2018juc}.
In particular, we use the result from Ref.~\cite{Archidiacono:2019wdp} which states that $\mathfrak{a}_4\xi^4<30\Mpcinv$ ($95\%$ C.L.), where $\xi$ is the ratio of the temperature of the DR bath to that of the CMB, which we fix at $\xi=0.5$.
This translates into a constraint of $\mathfrak{a}_4<480\Mpcinv$ from CMB and Lyman-$\alpha$ data, two orders of magnitude weajer than our forecasted 21-cm constraints.

Additionally, we now forecast the 21-cm fluctuation constraints under the other two foreground assumptions described in Sec.~\ref{sec:fluctuations}, for each of the models.
For optimistic foregrounds we find that HERA could measure up to $m_{\rm WDM}= 24$ keV, $m_{\rm FDM}=8.4\times 10^{-20}$ eV, or down to ETHOS couplings of $\mathfrak {a}_4=0.3 \Mpcinv$.
If, on the other hand, we assume pessimistic foregrounds, HERA could probe $m_{\rm WDM}= 10$ keV, $m_{\rm FDM}=3.0\times 10^{-20}$ eV, or down to ETHOS couplings of $\mathfrak {a}_4=2 \Mpcinv$, which is comparable to (albeit slightly better than) the results obtained with the global signal.
In all cases the first 2 PCs carry almost all the information, both for fluctuations and global signal, as the expected error-bars grow dramatically after that (see Fig.~\ref{fig:PCerrors}).

There are a few words of caution we would like to say before finishing this section.
First, in our simplified model the changes over the fiducial power spectrum are redshift-independent, whereas in reality for different DM models that is not the case, as there can be regeneration of power~\cite{Boehm:2003xr}.
Second, throughout this work we have assumed a sharp-$k$ window function, which fits well the WDM and FDM cases~\cite{Schneider:2018xba}, but is known to fail for DM-DR interactions~\cite{Sameie:2018juk}.
Once 21-cm data is gathered, and analyzed, these assumptions will be revisited.
Last, the stellar-formation history changes for different nCDM models~\cite{Bode:2000gq,Lovell:2017eec}, which would affect the inferred value of astrophysical parameters, such as $f_*$.
Here we are varying the astrophysical parameters independently of the DM properties, but a full analysis might yield stronger constraints.

\section{Conclusions}
\label{sec:conclu}

In this work we have studied how upcoming 21-cm measurements during cosmic dawn provide a powerful handle on the small-scale structure of our universe.
The galaxies that were able to cool gas into stars during cosmic dawn had masses $M\sim 10^6-10^8 M_\odot$, and thus were formed out of matter fluctuations with wavenumbers as large as $k\sim 100 \Mpcinv$.
Changing the matter power spectrum over those wavenumbers would, thus, alter the abundance of stars during cosmic dawn, affecting the 21-cm signal in an observable way.
By jointly varying astrophysical parameters and the matter power spectrum at different wavenumbers, we have shown that measurements of the 21-cm global signal, as well as its fluctuations, provide an indirect measurement of matter fluctuations for $k\lesssim 80\Mpcinv$, beyond the reach of other cosmic observables.

Broadly speaking, large-scale matter fluctuations with $k\lesssim40\Mpcinv$, act as a normalization of the number of stars.
These modes are, thus, highly degenerate with astrophysical parameters, such as the fraction $f_*$ of gas that forms stars, or the impact $B$ of the LW feedback.
Smaller scales, with $k\sim 40-80\Mpcinv$, alter the amount of stars with a different redshift behavior (see, e.g., Fig.~\ref{fig:dlogFcoll}), that allows us to more readily distinguish them from astrophysics.
The smallest scales ($k\gtrsim 80\Mpcinv$) will, however, affect tiny haloes that only formed stars at very high redshifts, and thus cannot be well measured.
This broad reasoning is validated by our detailed forecasts, as we find that the 21-cm global signal can measure the amplitude of the matter power spectrum integrated over $k = (40-80) \Mpcinv$ to $\sim 30\%$ level precision.
The 21-cm fluctuations will, moreover, improve this result by allowing us to probe two different $k$-bins over the same range, spanning $k = (40-60)$ and $(60-80) \Mpcinv$, to $\sim 10\%$ precision.

Nonetheless, different $k$-bins of matter fluctuations are highly degenerate.
To overcome this hurdle, we have performed a principal component analysis using both the 21-cm global signal and its fluctuations.
Therefore, the amplitude of the PCs found here can be jointly fit with astrophysical parameters, as opposed to varying it at each individual wavenumber $k$, still obtaining a nearly maximal amount of information.
We have found that the 21-cm global signal allows us to measure 2 principal components (PCs) with signal-to-noise ratios (SNR) larger than five.
The 21-cm fluctuations, on the other hand, allow for $\{3,4,8\}$ PCs to be measured under the assumption of pessimistic, moderate, and optimistic foregrounds.
We used these PCs to obtain model-agnostic constraints on the matter power spectrum, showing that they are mostly sensitive to wavenumbers $k\sim 40-80\Mpcinv$.
We projected several non-CDM models onto our PCs, finding that the 21-cm signal during cosmic dawn can improve the constraints on all of these models over other current cosmic probes, such as the Lyman-$\alpha$ forest. 

Our focus in this paper was to probe the small-scale matter power spectrum.
Nevertheless, it is worth noting that our results extend to inflationary features in the primordial power spectrum, which can dramatically increase or decrease the power beyond CMB scales.
Additionally, we have shown that altering the power spectrum at scales as small as $k\sim 50\Mpcinv$ produces effects similar to changing some astrophysical parameters, such as $f_*$, so careful modeling of the smallest scales is required even if one was only interested in understanding astrophysics.
Finally, it is possible to detect haloes with masses $M\sim 10^6-10^8M_\odot$ with other probes, such as strong lensing~\cite{Dalal:2001fq,Vegetti_2010_2,Vegetti_2012,Ritondale:2018cvp,2014MNRAS.442.2017V,Cyr-Racine:2015jwa,Hezaveh:2016ltk,Rivero:2017mao, Rivero:2018bcd, Gilman:2019vca,DiazRivero:2019hxf}, CMB weak lensing~\cite{Nguyen:2017zqu}, or Milky-Way satellites~\cite{Jethwa:2016gra,2016MNRAS.457.3817S,Nadler:2018iux,Banik:2019smi,Nadler:2019zrb}.
Nonetheless, low-redshift probes measure the abundance of small haloes in the highly nonlinear regime, where any deviation due to dark-matter physics has been highly processed and can be washed away.
We are able to probe the abundance of small haloes first becoming nonlinear, and collapsing, with the 21-cm line.
Thus, any deviation from CDM will appear more pristine in cosmic-dawn studies.

In summary, we will gain a wealth of information about cosmology from the 21-cm line. 
Despite large astrophysical unknowns, we have shown that we can use measurements of the 21-cm global-signal and fluctuations to probe the matter power spectrum at currently unobserved scales.
This will open up a new window into the nature of dark matter and the physics of cosmic inflation.

\acknowledgements
It is our pleasure to thank Prateek Agrawal and Adam Lidz for discussions.
CD and JBM were supported by NSF grant AST-1813694. 
Some computations in this paper were run on the FASRC Odyssey cluster supported by the FAS Division of Science Research Computing Group at Harvard University.

\bibliography{pk_21cm}

\appendix

\end{document}